\documentclass[10pt,journal,compsoc]{IEEEtran}

\ifCLASSOPTIONcompsoc
  \usepackage[nocompress]{cite}
\else
  \usepackage{cite}
\fi

\ifCLASSINFOpdf
\else
\fi

\usepackage{cite}
\usepackage{amsmath,amssymb,amsfonts}
\usepackage{algorithmic}
\usepackage{graphicx}
\usepackage{textcomp}
\usepackage{xcolor}
\usepackage{colortbl}
\usepackage{fancyhdr}

\usepackage[hidelinks]{hyperref}

\begin{document}
\title{E-BATCH: Energy-Efficient and High-Throughput RNN Batching}
\setlength{\abovedisplayskip}{2pt}
\setlength{\belowdisplayskip}{2pt}

\author{Franyell~Silfa,
        Jose Maria~Arnau,
        and~Antonio~Gonz\'{a}lez%
        \break
         \thanks~Universitat Polit\`{e}cnica de Catalunya (UPC), Barcelona, Spain

        }

\markboth{}%
{Silfa \MakeLowercase{\textit{et al.}}: E-BATCH: Energy-Efficient and High-Throughput RNN Batching}

\IEEEtitleabstractindextext{%
\begin{abstract}

Recurrent Neural Network (RNN) inference exhibits low hardware utilization due to the strict data dependencies across time-steps. Batching multiple requests can increase throughput. However, RNN batching requires a large amount of padding since the batched input sequences may largely differ in length. Schemes that dynamically update the batch every few time-steps avoid padding. However, they require executing different RNN layers in a short timespan, decreasing energy efficiency. Hence,
we propose E-BATCH, a low-latency and energy-efficient batching scheme tailored to RNN accelerators. It consists of a runtime system and effective hardware support. The runtime concatenates multiple sequences to create large batches, resulting in substantial energy savings. Furthermore, the accelerator notifies it when the evaluation of a sequence is done, so that a new sequence can be immediately added to a batch, thus largely reducing the amount of padding. E-BATCH dynamically controls the number of time-steps evaluated per batch to achieve the best trade-off between latency and energy efficiency for the given hardware platform.
We evaluate E-BATCH on top of  E-PUR and TPU. In E-PUR, E-BATCH improves throughput by 1.8x and energy-efficiency by 3.6x, whereas in TPU,  it improves throughput by 2.1x and energy-efficiency by 1.6x, over the state-of-the-art.

\end{abstract}

\begin{IEEEkeywords}
Machine Learning, Accelerators, Long Short Term Memory, Recurrent Neural Network, Batching.
\end{IEEEkeywords}}

\maketitle

\IEEEdisplaynontitleabstractindextext

\IEEEpeerreviewmaketitle

\section{Introduction}\label{s:introduction}
Recurrent Neural Networks (RNNs) are a key technology for sequence-to-sequence applications such as machine translation~\cite{britzGLL17} and speech recognition~\cite{deepspeech2}.
Their connections include feedback loops that allow them to remember information from previous executions and handle input and output sequences of variable length (i.e., number of time-steps). Modern RNNs feature a large number of parameters and, hence, their memory requirements are in the order of tens and even hundreds of megabytes~\cite{fsilfa2018epur}. Also, they impose severe data dependencies, and as a result, RNN inference exhibits a limited amount of parallelism. Not surprisingly, state-of-the-art systems such as TPU~\cite{jouppi2017TPU} or Brainwave~\cite{brainwave2018} exhibit a low
resource utilization for RNN inference: 18\% and 3.5\% respectively. On the other hand, other accelerators such as E-PUR~\cite{fsilfa2018epur} achieve nearly 100\% utilization but are tailored to mobile environments. GPUs/CPUs using state-of-the-art RNN libraries also exhibit extremely low resource utilization. For example, the high-performance library cuDNN~\cite{cudnn} shows an average utilization of 13.5\% on an NVIDIA Titan V GPU for RNN inference.

Servers handling multiple requests (i.e., cloud services) from a large number of edge devices, such as smartphones, employ batching to increase parallelism and throughput. During inference, batching merges several requests and feeds them to the system at the same time, so that all their computations are done in parallel. Therefore, the high energy cost of accessing the model parameters is shared by all the requests in a batch.
Note that batching works best when the batched requests are identical in length, i.e., their number of time-steps are the same. However, this is particularly difficult in RNNs since their input sequences usually have different number of time-steps. For instance, Deepspeech2~\cite{deepspeech2} has input sequences with several time-steps ranging from 60 to 1700 (Librispeech test set).

State-of-the-art deep learning systems~\cite{tensorflow2016, pytorch2017} handle this issue by padding the batched sequences such that their number of time-steps are identical to the number of time-steps of the longest sequence. The main drawback of this approach is that the latency of all the batched sequences increases since their evaluation cannot be completed until the longest sequence has been evaluated. Besides, energy is wasted performing computations on the extra added time-steps. Our experiments on E-PUR, an RNN accelerator, show that 30.2\% of the energy consumption is due to padding, whereas the latency overhead is 28.5\% on average for a set of RNNs.
 
For Deep RNN models, weight reuse is severely affected by the number of requests in a batch and the number of time-steps in the batched sequences. The reason is that, in order to evaluate a new layer of an RNN model, the weights are first brought to on-chip memory, hence evicting the weights of the previous layer. Henceforth, creating batches with short sequences incurs a large amount of weight swapping, and, as a consequence, energy consumption increases.
This issue is evident in batching strategies such as Cellular Batching~\cite{gao2018cellular}, where batches with short sequences are created, and not surprisingly, it is inefficient energy-wise. For instance, cellular batching on top of E-PUR consumes, on average, 4.5x more energy per request than sequence padding for DeepSpeech~\cite{deepspeech2}.

 Motivated by the inefficiencies of current batching schemes, we propose E-BATCH, an RNN batching scheme that improves energy efficiency by avoiding padding and by increasing the temporal and spatial locality of the weights. In E-BATCH, sequences are allowed to join a batch while it is being evaluated. Also, when a new batch is created, all the available requests are distributed among all the hardware \textit{processing lanes} (i.e., a processing element that can evaluate one or more sequences sequentially). More specifically, for a system with \textit{n} processing lanes, we partition all the available requests among processing lanes in a way that the difference between the total number of time-steps evaluated by each lane is minimal. In case that the number of available requests is larger than the number of lanes, several requests are assigned to a given processing lane and are processed sequentially. Also, when the number of requests is too low, the system waits for some time before starting a new batch to increase its size. Furthermore, while a batch is being evaluated, specific new requests are allowed to join the batch to increase its efficiency. Note that assigning more than one request to a processing lane increases the length of the batched sequences.
Furthermore, to meet Service-Level-Agreement~(SLA), we limit the maximum number of time-steps in a given processing lane. Our experiments show that, on average, E-BATCH improves energy efficiency and throughput by 2.6x and 1.9x, respectively.  For the rest of the paper, we refer to processing lanes simply as lanes.

To summarize, in this paper, we focus on batching for RNN inference. More specifically, batching for LSTM and GRU networks on DNN accelerators. Its main contributions are the following:

\begin{itemize}
    \item
We analyze the trade-off between energy and latency while batching RNN sequences.  Also, we identify the excessive padding and the poor weight locality as the primary sources of inefficiencies in current RNN batching approaches.  
    \item We propose a novel batching scheme that largely improves temporal and spatial locality of the weights and minimizes padding, which results in significant 
    throughput and energy improvements.
    \item We implement our scheme on top of E-PUR and TPU, two state-of-the-art accelerators for RNNs. Our system improves energy efficiency by 3.6x/2.1x and throughput by 1.8x/1.6x  for E-PUR/TPU, respectively, compared to a state-of-the-art RNN batching strategy.
\end{itemize}

\section{Background}\label{s:background}

 \begin{figure}[t!]
	\centering
	\includegraphics[width=3.375in]{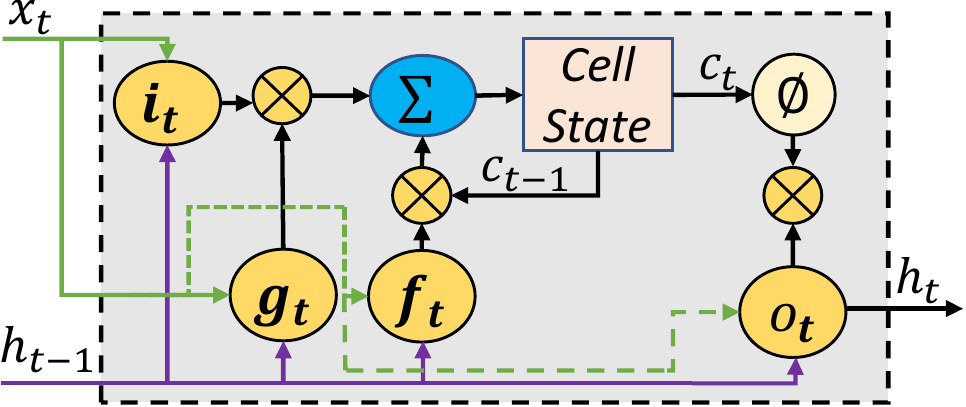}
	\caption{Structure of an LSTM cell.}
	\label{f:lstm_cell}
\end{figure}

\subsection{Recurrent Neural Networks}\label{s:rnn_networks}
 Recurrent Neural Networks are a state-of-the-art machine learning algorithm for sequence-to-sequence problems
 such as machine translation~\cite{britzGLL17}, video captioning~\cite{vinyalsTBE16} and speech recognition~\cite{deepspeech2, miao2015eesen}.
 One of the main characteristics of RNNs is that they include feedback loops that allow them to use past information from previously executed time-steps. Furthermore,
 RNNs are evaluated recurrently for each time-step of the input sequence. Therefore, they can handle input and output sequences of variable length. These features make them an extremely effective framework for sequence-to-sequence problems for which they typically outperform conventional DNNs and CNNs.

  Basic RNN (i.e., vanilla RNN) cannot capture long dependencies in the input sequence because information tends to dilute over time.
 For this reason, the Long Short Term Memory (LSTM)~\cite{hochreiter1997long} and Gated Recurrent Unit (GRU)~\cite{cho14Gru} architectures were proposed.  
 Since they can exploit long and short term dependencies, they represent the most successfully and commonly used RNN architectures nowadays.
 
 In an RNN, an input sequence (e.g. an audio frame) is composed of \textit{N} time-steps, i.e.  $X = [x_1, x_2, ..., x_N]$. In an LSTM or GRU network,  time-steps are processed sequentially in each LSTM or GRU cell, from $x_1$ to $x_n$. Note that the number of time-steps is normally different for each input sequence that is evaluated.

\begin{figure}[t!]
	\centering
	\begin{align}
	i_t = \sigma(W_{ix} x_t + W_{ih} h_{t-1}  + b_i)
	\label{e:input_gate}
	\end{align}
	\begin{align}
	f_t = \sigma(W_{fx} x_t + W_{fh} h_{t-1}  + b_f)
	\label{e:forget_gate}
	\end{align}
	\begin{align}
	g_t = \phi(W_{gx} x_t + W_{gh} h_{t-1} + b_g)
	\label{e:update_gate}
	\end{align}
	\begin{align}
	c_t = f_t \odot c_{t-1} + i_t \odot g_t
	\label{e:cell_state}
	\end{align}
	\begin{align}
	o_t = \sigma(W_{ox} x_t + W_{oh} h_{t-1}  + b_o)
	\label{e:output_gate}
	\end{align}
	\begin{align}
	h_t = o_t \odot \phi(c_t)
	\label{e:cell_output}
	\end{align}
	\caption{Computations of an LSTM cell. $\odot$, $\phi$, and $\sigma$ denote element-wise multiplication,
		hyperbolic tangent and sigmoid function respectively.}
	\label{f:lstm_equations}
\end{figure}

Regarding Deep RNNs, they
are created by stacking together several layers. Each of these layers contains an LSTM or GRU cell. 
Furthermore, each layer can be unidirectional or bidirectional. Bidirectional layers use past and future information to
make inference. On the contrary, unidirectional layers only use past information.

 \subsubsection{Basic structure of an LSTM Cell}\label{s:lstm_cell}

Figure~\ref{f:lstm_cell} shows the structure of an LSTM cell. The principal component of this cell is the cell state (${c_t}$, in Equation~\ref{e:cell_state}), which is 
used to store information. The cell state is computed as a function of four different gates in charge of modulating the amount of information added or deleted from it on each time-step. The input gate 
 ($i_t$, in Equation~\ref{e:input_gate}) modulates how much of the input information is added to the cell state. In contrast,  
the forget gate ($f_t$, in Equation~\ref{e:forget_gate}) controls how much information is removed from the cell state. 
The updater gate ($g_t$, in Equation~\ref{e:update_gate}) modulates the amount of information 
that is considered as a candidate to update the cell state. Finally, the output gate ($o_t$, in Equation~\ref{e:output_gate}) 
controls how much information from the cell state is exposed as the cell output ($h_t$, in Equation~\ref{e:cell_output}).

The computations performed by an LSTM cell are shown in Figure~\ref{f:lstm_equations}. For each gate, there are two types of inputs:
one is the current time-step($x_t$) and the other is the previous output ($h_{t-1}$). In order to compute each gate's output, two matrix-vector multiplications are performed: one between the input $x_t$ and $W_x$, and another between $h_{t-1}$ and $W_h$. After these multiplications, an activation function is applied, which is typically a sigmoid or hyperbolic tangent. The output of each gate is a vector, and for the sake of simplicity, we call its elements \textit{neurons}. Most of the execution time in an LSTM cell is due to the computation of the matrix-vector multiplication. In contrast, most of the energy consumption is due to accessing the weight matrices~\cite{fsilfa2018epur}.

Regarding GRU Cells, they work similarly to LSTM cells, but
they do not include a cell state. A GRU cell is composed of two gates: the update gate ($z_t$) and the reset gate ($r_t$).
The update gate ($z_t$)  modulates how much information from the previous output ($h_{t-1}$) will be carried over the current output ($h_t$). On the other hand, the reset gate ($r_t$) controls how much information from the previous
output is removed. The computations performed by a GRU cell are similar to the equations shown in Figure~\ref{f:lstm_equations}, so we omit them for the sake of brevity. In the rest of the paper, we refer to both LSTM and GRU cells as RNN cells.

\begin{figure}[t!]
	\centering
	\includegraphics[width=3.375in]{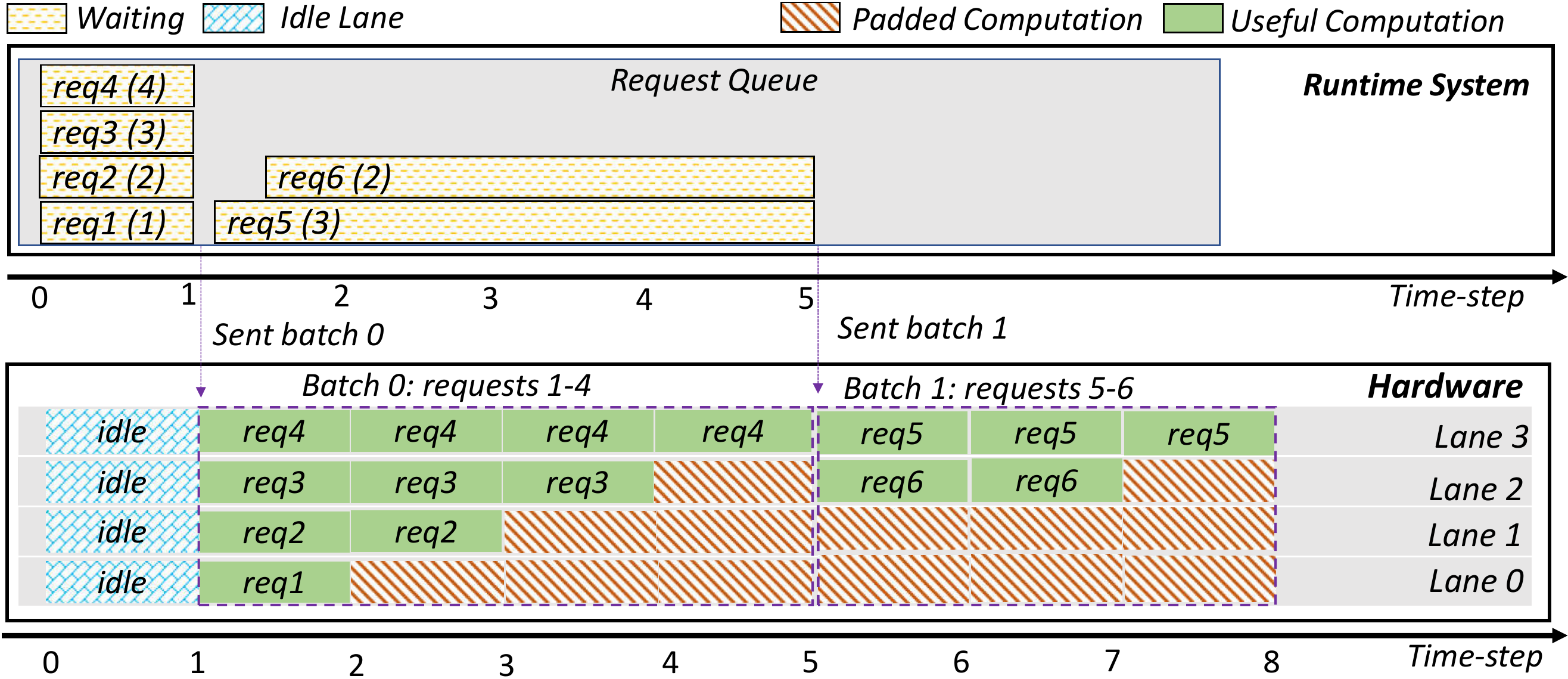}
	\vspace*{-2mm}
	\caption{Sequence Padding. Requests are shown in the request queue from their arrival time until they are dispatched to the hardware for evaluation. The number inside the parenthesis next to each queued request is the number of time-steps for that request. The batch size is 4.   }
	\label{f:padding_example}
\end{figure}

 \subsection{Sequence Batching}\label{s:batching}
Sequence Batching (we will also refer to it as batching for short) is a well-known technique commonly used to increase throughput. Inference machine learning systems handling multiple requests (e.g., data centers providing service to many users) batch various requests so that their computations are done in parallel. In this context, a batch is a set of one or more requests (i.e., input sequences). The number of batched requests (i.e., batch size) is usually limited by the amount of hardware resources available in the system (e.g., number of processing lanes). Arriving requests are grouped into batches of size $N$, and batches are evaluated sequentially. 
 
 Usually, once a batch of requests is sent to the hardware for evaluation, each of the batched requests is assigned to a processing lane. Then, the evaluation of any of those requests is not completed until all of them are computed. Henceforth, batching tends to work best when the batched requests have the same length (i.e., an equal number of time-steps among input sequences). The reason is that all the processing lanes are fully utilized, and none of the batched requests will have to wait for others to complete. However, this is an issue in RNNs since the number of time-steps on each input sequence is typically different. Henceforth, to mitigate this problem, the following strategies are commonly employed.

\begin{figure}[t!]
	\centering
	\includegraphics[width=3.375in]{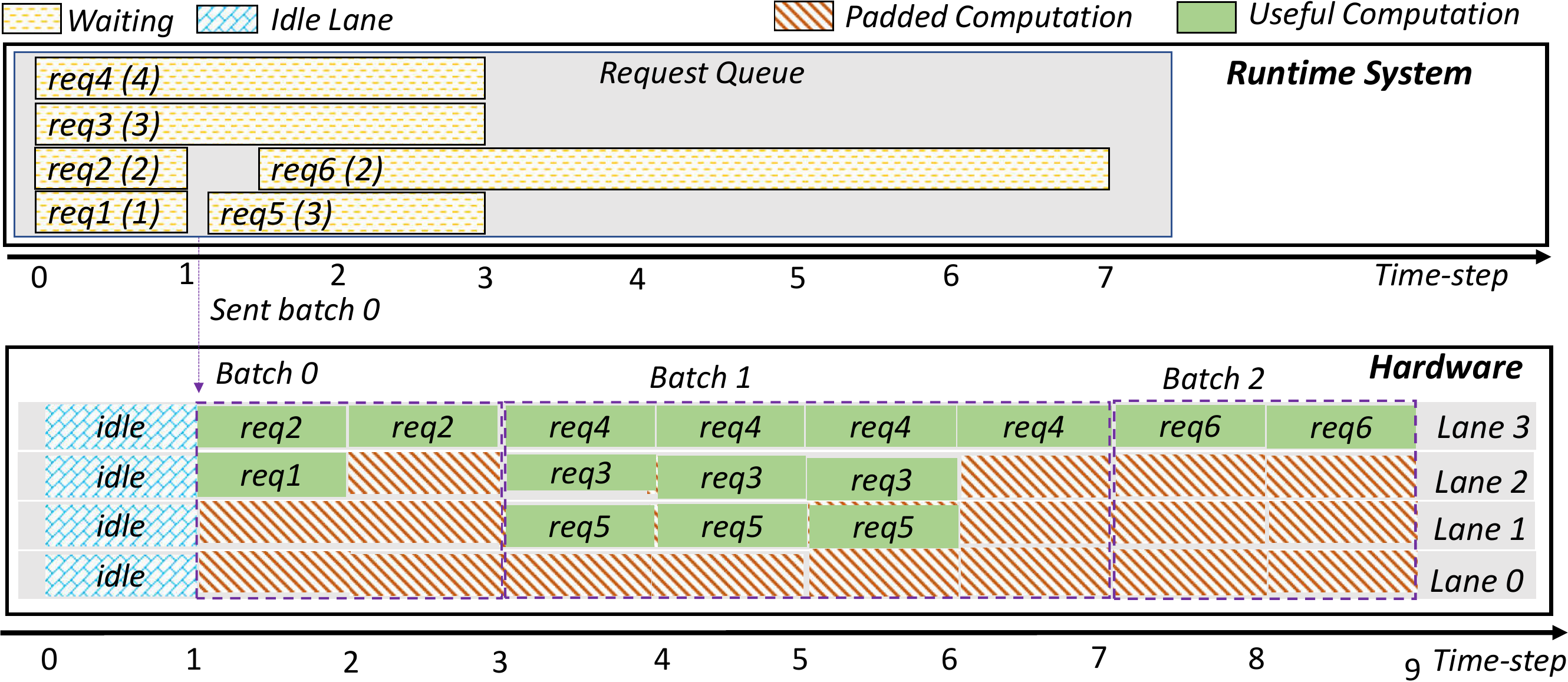}
	\vspace*{-2mm}
	\caption{
Sequence Bucketing. For this example, the maximum difference in time-steps for requests that are batched together is 1.  }
	\label{f:bucketing_example}
\end{figure}

 \subsubsection{Sequence Padding}\label{s:batching_padding} 

 Sequence Padding is used in systems such as TensorFlow~\cite{tensorflow2016} and PyTorch~\cite{pytorch2017} to handle sequences with different number of time-steps when batching is performed for RNNs. In this case, the sequence with the largest amount of time-steps (i.e., \textit{m}) in a batch is found. Then, for each of the other batched sequences, their number of time-steps is increased to match the maximum length \textit{m}. These extra added time-steps
 are filled with zeros, and the larger the difference among batched input sequences, the larger the amount of useless computations. As an example consider the batches created in Figure~\ref{f:padding_example}, where \textit{requests 1-6} have 1, 2, 3, 4, 3, and 2 time-steps, respectively. Hence, in order to create a batch of size 4, \textit{requests 1-3} are padded. Note that \textit{request 1} completes its execution long before \textit{request 4} (i.e. the longest request in the batch), but it is not returned to the user until \textit{request 4} is finished. Furthermore, although \textit{request 5} is available when \textit{request 1} is already finished, its computation cannot start until the whole batch is computed.

 \subsubsection{Sequence Bucketing}\label{s:batching_bucketting} 
Sequence Bucketing~\cite{bucketing2016} is an optimization technique that is 
 implemented on top of sequence padding in systems such as MXNet~\cite{cheMxnet} and TensorFlow. Its target is to reduce padding, hence decreasing the amount of wasted computations. In order to accomplish this, different sequences are clustered together into a logical group, a.k.a. \textit{bucket}, based on a given heuristic. One commonly used heuristic, shown in Figure~\ref{f:bucketing_example}, is to assign sequences of similar length to a given 
 \textit{bucket}. In this regard, the similarity is defined as the maximum difference in time-steps among all the sequences in any given \textit{bucket}. Also, it is constrained to be below a given threshold (i.e., the \textit{bucket width}). 
 
When a batch is created during sequence bucketing, only sequences from the same \textit{bucket} can be batched together. Note that some of the sequences in a given batch may still require some padding. The maximum amount of time-steps padded is the \textit{bucket width}. For instance, consider the example in Figure~\ref{f:bucketing_example}, assuming that \textit{requests 1-6} have 1, 2, 4, 5, 3, and 2 time-steps respectively, and the \textit{bucket width} is one. \textit{Requests 1, 2, and 6} are assigned to a \textit{bucket} whereas \textit{request 3-5} are assigned to another \textit{bucket}. Then, if batches are created using a batch size of 4, \textit{requests 1-2} will be batched together. Similarly, \textit{requests 3-5} will go into the same batch. Note that, although \textit{request 6} is available when \textit{requests 3-5} are batched, it is not included since it belongs to a different \textit{bucket}. Analogous to sequence padding, batches are evaluated sequentially, and new requests are not allowed to join a given batch while it is being evaluated in the hardware. In the rest of the paper, we refer to this optimization as bucketing.

 \subsubsection{Cellular Batching}\label{s:batching_cellular}
\begin{figure}[t!]
	\centering
	\includegraphics[width=3.375in]{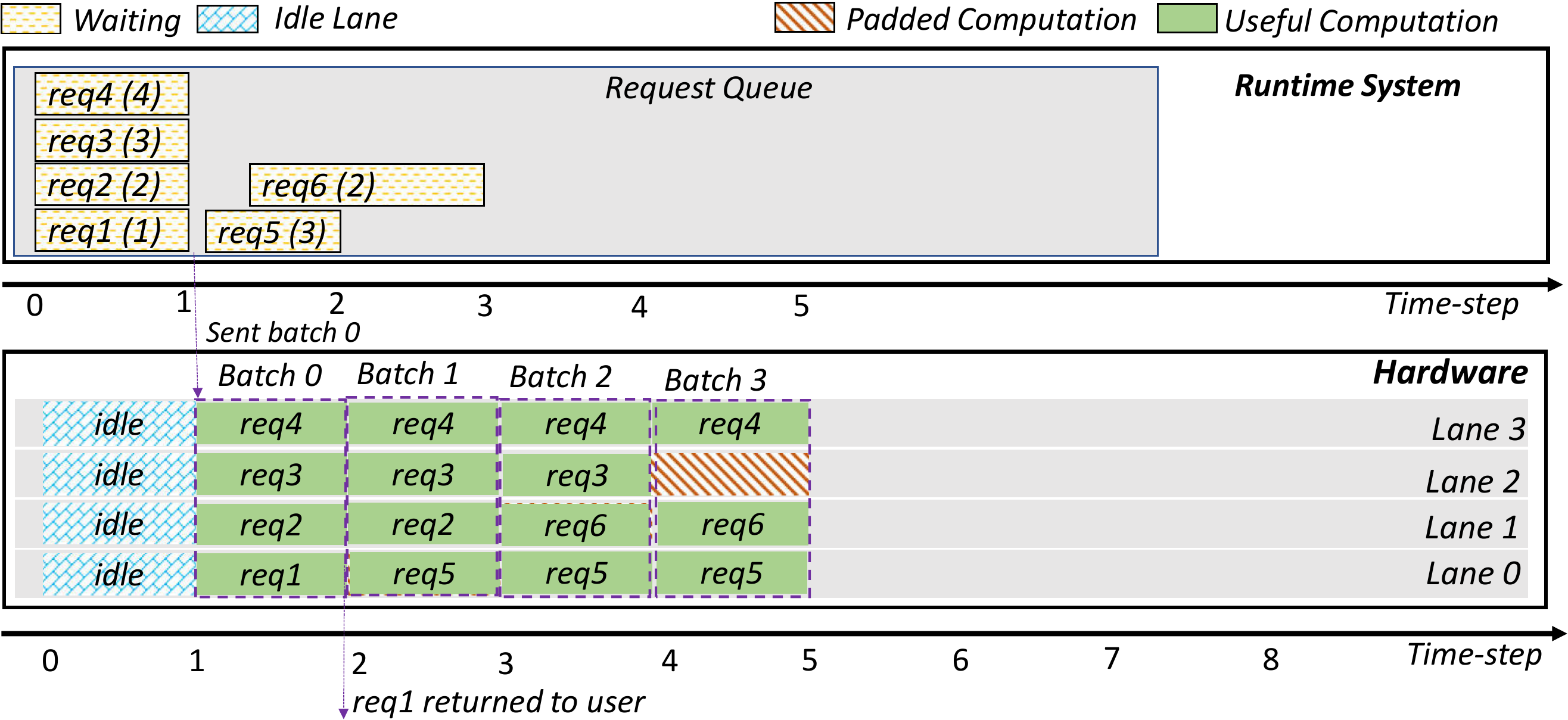}
	\vspace*{-2mm}
	\caption{Cellular Batching. }
	\label{f:cellular_batching_example}
\end{figure}

 Cellular Batching is a recently proposed technique for RNN batching that focuses on batching requests at the granularity of cells (i.e., some time-steps) instead of whole sequences~\cite{gao2018cellular}. Unlike sequence padding and bucketing, in cellular batching new requests are allowed to join a batch whose execution is ongoing. Also, once a request is evaluated, it can be returned to the user immediately. Figure~\ref{f:cellular_batching_example} shows a simple example of how requests are scheduled in cellular batching. For this example, we assume a cell represents one time-step, a batch size of 4, and an LSTM model of one layer. In this example, the first time-steps of \textit{requests 1-4} are batched together and sent to the hardware for evaluation. Once the batch is evaluated, \textit{request 1} is completed and returned to the user. After this, a new batch is created using the second time-step from \textit{requests 2-4} and time-step 1 from \textit{request 5}. Once this batch is evaluated, \textit{request 2} is completed and sent to the user. Then, batching and evaluation process continues for the remaining time-steps of \textit{requests 3-6} until all of them are evaluated. Since batches are created using a fine granularity, new requests can start execution as soon as the processing hardware becomes available, and completed requests are sent to the user immediately. Hence, the time a request waits in the queue is reduced. As a result, the average latency and throughput of the system are improved.

\section{Sources of batching inefficiencies in RNNs}\label{s:batching_inefficiencies}

Both Bucketing and Cellular batching provide an improvement over sequence padding by reducing the amount of wasted computations. Cellular batching
also improves latency by batching requests using a finer granularity. However, these solutions do not take into account energy consumption and
the spatial and temporal locality of the weights for large RNN models (i.e., more than one layer), which is typically exploited in RNN accelerators. In this section, we identify the sources of inefficiencies in RNN batching and present a detailed analysis.

\begin{figure}[t!]
	\centering
	\includegraphics[width=3.375in]{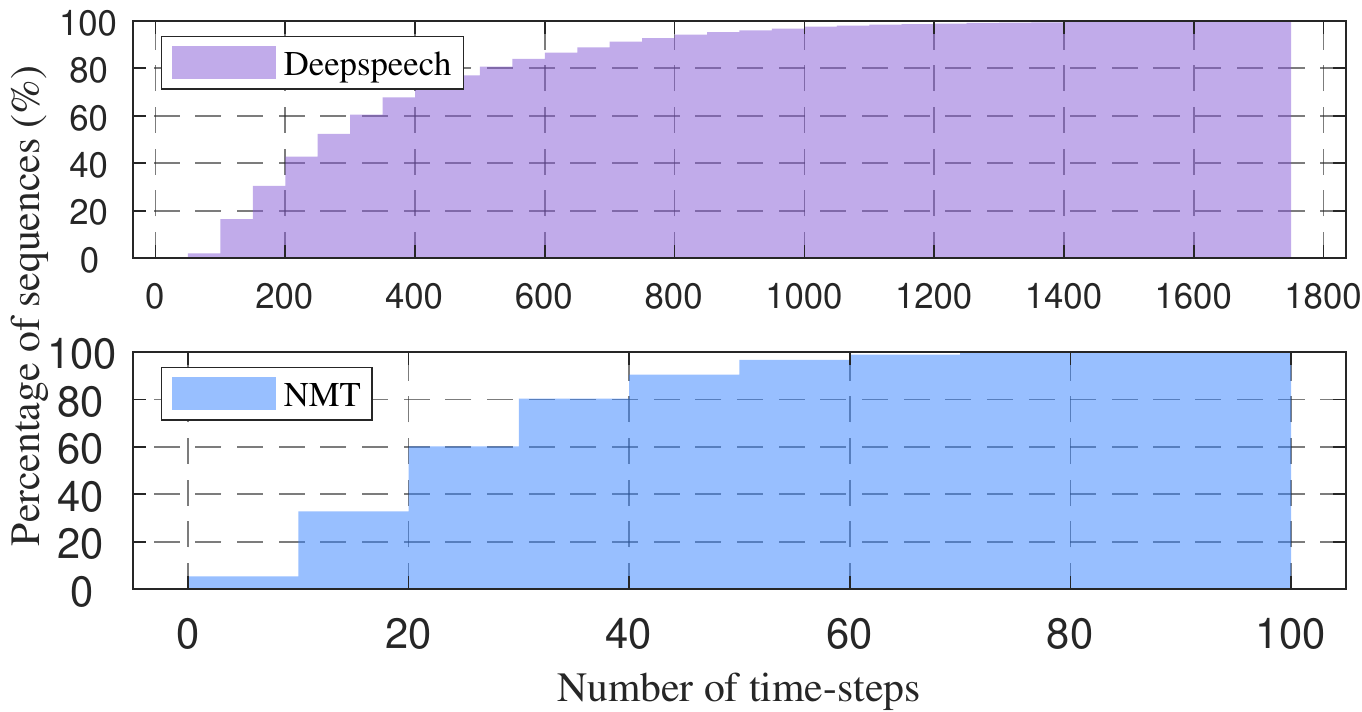}
	\vspace*{-2mm}
	\caption{Time-steps distribution for Deepspeech and NMT. The number of time-steps ranges from 10 to a few hundred.}
	\label{f:input_variabilty}
\end{figure}

\subsection{Number of Times-steps Variability}\label{s:sequence_variability}

Figure~\ref{f:input_variabilty} shows the cumulative distribution for the number of time-steps in the input sequences of two popular RNN models. As it can be seen, 
 both models have a wide range of sizes in the length of their input sequences. This variability severely affects batching systems that employ strategies such as sequence padding or bucketing since many useless computations are introduced. For instance, we have seen in our experiments that nearly 40\% of the calculations evaluated by the hardware are unnecessary for sequence padding. On the contrary, when bucketing is applied, the number of wasteful computations evaluated decreases to nearly 5\%, since the number of time-steps among sequences that are batched together are quite similar. Regarding cellular batching, less than one percent of the computations executed are useless since batches are created and evaluated at a fine granularity (i.e., five time-steps).

\begin{figure}[t!]
	\centering
	\includegraphics[width=3.375in]{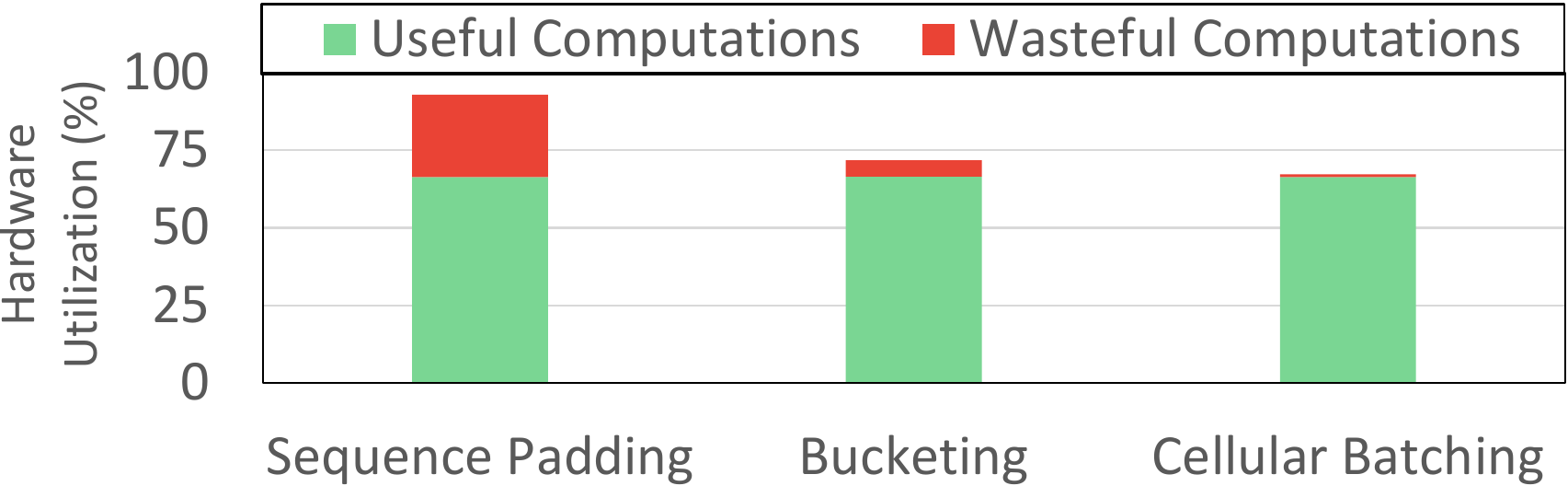}
	\vspace*{-2mm}
	\caption{Percentage of hardware utilization for useful and wasteful computations for Deepspeech.}
	\label{f:padding_breakdown}
\end{figure}

\subsection{Low Hardware Resource Utilization}\label{s:resource_utilization}

There are two leading causes of low hardware utilization of RNNs systems. One is due to padded sequences, whereas the other is due to a limited amount of requests.

Figure~\ref{f:padding_breakdown} shows a breakdown for the percentage of hardware utilization for useful and wasteful computations on a system under a moderate workload (i.e., 1000 requests per second ).
As it can be seen, when employing sequence padding around 26\% of the hardware utilization is due to wasteful computations. On the contrary, when bucketing is employed, only 5.2\% of the hardware is utilized for wasteful calculations.

Some wasteful computations are performed by lanes that complete the evaluation of requests assigned to them early. 
 For these cases, we could reduce the number of wasteful computations by sending available requests to those lanes, since there are no data dependencies among requests. Note that, typically when a batch is being evaluated in the hardware, to increase weight reuse, the same model parameters (i.e., weights) 
 are used by each processing lane. Therefore, time-steps from an available request can only be sent to a lane when its weights are the same as the weights being used to compute the current set of batched requests. We provide more details about this issue in Section~\ref{s:technique}.

On the other hand, when the number of requests waiting to be processed is not enough to fill the available lanes, 
batches of small sizes are created. For this reason, some lanes will perform padded computations. Besides and more importantly, weights are swapped more frequently (i.e., for each new batch and layer evaluated), 
hence decreasing weight reuse and dramatically increasing energy consumption. For instance, for the model evaluated in Figure~\ref{f:padding_breakdown}, 
when the workload is low (i.e., number of requests is smaller than the batch size), waiting for more new requests to arrive to create batches with a
 more considerable amount of requests, decreases energy consumption per request processed by 2.3x on average.

\begin{figure}[t!]
	\centering
	\includegraphics[width=3.375in]{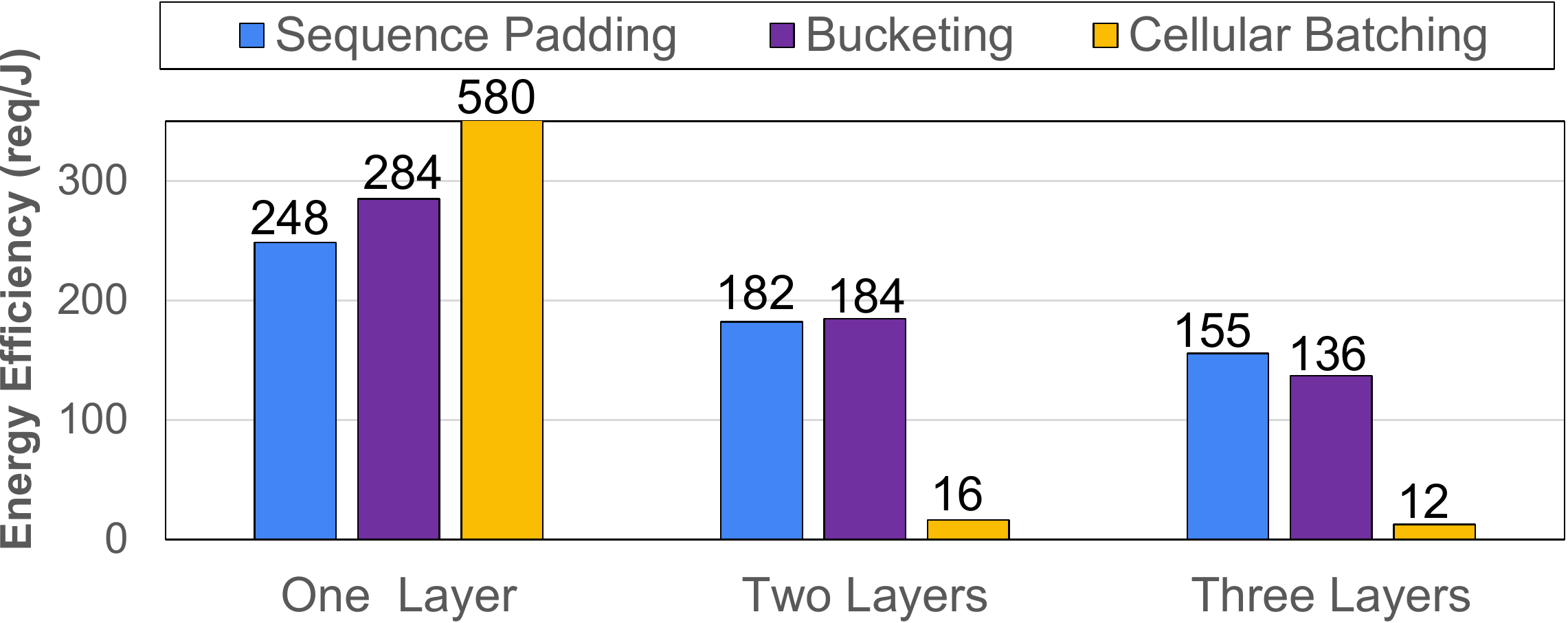}
	\vspace*{-2mm}
	\caption{Energy Efficiency of sequence padding and cellular batching for the GNMT~\cite{wu2016google} model on E-PUR. The system load is 100 request per second and a batch size of 64.}
	\label{f:energy_cell_pad_buck}
\end{figure}

\subsection{Poor Weight Locality}\label{s:weight_locality}

For RNN inference, one of the primary sources of energy consumption is the memory accesses to fetch the weights~\cite{fsilfa2018epur}, accounting for up to 80\% of overall energy consumption. Therefore, hardware accelerators for RNN include local on-chip memories to increase weight reuse. Not surprisingly, when batching RNN sequences,
this keeps being the dominant factor. 

For Deep RNNs, on each batch evaluation, the weights must be fetched for each layer. Note that while evaluating an RNN layer, due to data dependencies, all the time-steps of a group of sequences in a batch are usually computed before proceeding with deeper layers. Hence, depending on the number of time-steps of the batched sequences, the overhead of accessing the weights will be more or less severe. This is particularly problematic when evaluating cellular batching on accelerators because input sequences are split into multiple batches. Moreover, batches, where the sequences on them have a small number of time-steps (e.g., 5), are created. Therefore, for RNN models with more than one layer, weights are fetched multiple times for each sequence on average.

\begin{figure}[t!]
	\centering
	\includegraphics[width=3.375in]{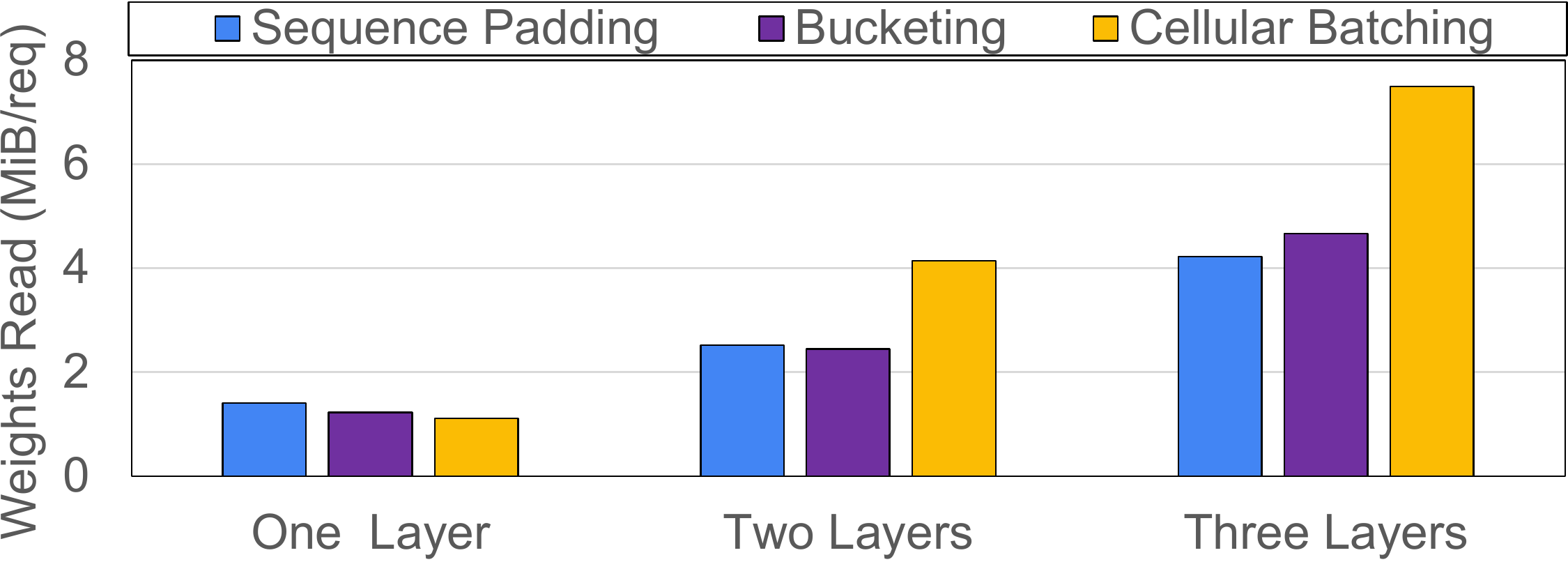}
	\vspace*{-2mm}
	\caption{Number of weight reads per request for sequence padding and cellular batching.}
	\label{f:bytes_read_cell_pad}
\end{figure}

Figure~\ref{f:energy_cell_pad_buck} shows the average energy consumed per request for each of the three previous batching strategies, evaluated on E-PUR~\cite{fsilfa2018epur}. As it can be seen, when 
evaluating 
an RNN model with only one layer, cellular batching is highly efficient since it avoids wasteful computations. Also, because it decreases waiting time, static energy consumption is improved. 
Note that for one layer model, weights are only loaded into on-chip memory only once.
Despite cellular batching being highly efficient for a one-layer model, it is very inefficient for deeper models. As shown in Figure~\ref{f:energy_cell_pad_buck}, when the model has more than one layer, the energy efficiency of padding and bucketing is 11x compared to cellular batching. The main issue with cellular batching for deep RNN models is that weights are swapped too frequently from on-chip memory. More specifically, batches with a small number of requests are evaluated, which decreases weight reuse. As illustrated in Figure~\ref{f:bytes_read_cell_pad}, in the case of cellular batching,
 increasing the number of layers increases the amount of bytes read per request rapidly. However, for bucketing and sequence padding, the increment in weight reads per request is less severe. Note that, for the three-layer model, sequence padding has better energy efficiency. The reason is that because of the\textit{bucket size} constraint, batches with a small number of requests are created for sequence padding. 

One approach to increase weight reuse is to use large batch sizes so that more requests are batched together. However, increasing the
batch size will also increment the amount of lanes needed to evaluate them. Another approach is to create batches in which several requests are concatenated before being sent to a processing lane, where they are evaluated sequentially. The main drawback of this approach is that it increases the number of time-steps evaluated sequentially per processing lane for a given batch, which increases the average latency. However, significant improvements in energy efficiency can be achieved (as shown later in Section~\ref{s:results}).

To improve energy efficiency in RNN batching, we develop E-Batch, a novel batching scheme that allows RNN sequences to join a batch while it is being executed dynamically. Also, our proposal strives to create batches with a low amount of padding by employing a multi-way greedy partition algorithm. Furthermore, to increase weight reuse, E-Batch tries to evaluate a large number of time-steps in each lane. Finally, to improve resource utilization, E-Batch implements a timeout so that more requests are batched together during intervals when the number of requests arriving at the system is low.

We present the details of E-Batch in the next sections of the paper. We implemented our proposal on top of E-PUR and TPU, two state-of-the-art RNN accelerators. First, a brief description of the overall hardware baseline is presented. Next, we detail the extensions added to the baseline to support batching. Then, E-Batch is described. Finally, we discuss the runtime and hardware support required by our proposal.

\begin{figure}[t!]
	\centering
	\includegraphics[width=3.375in]{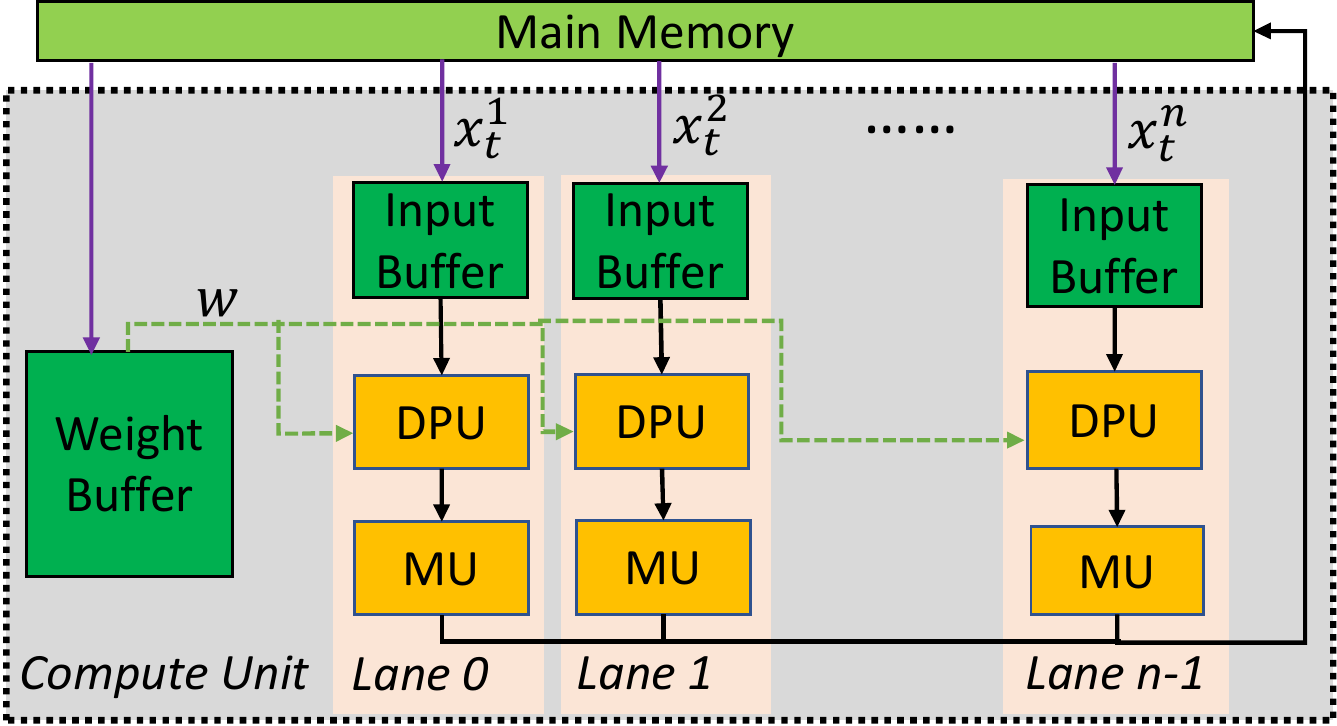}
	\caption{Architecture of a Compute Unit (CU) in E-PUR. In order to support batching, several processing lanes are included.}
	\label{f:computational_unit}
\end{figure}
\section{RNN Accelerator}\label{s:hardware}

The proposed RNN batching scheme, described in Section~\ref{s:technique}, can be implemented on top of any RNN accelerator. We first implement it on top of E-PUR, an energy-efficient accelerator for RNNs~\cite{fsilfa2018epur}. E-PUR consists of four computational units dedicated to evaluate the four gates in an LSTM cell. Moreover, it includes several on-chip memories to store the weights and intermediate results. In this work, we extend E-PUR to support batching. The next subsections provide further details on E-PUR's architecture and describe the required modifications. 

To illustrate the general applicability of the proposed batching technique, we also implement it on top of a TPU-like architecture and provide experimental results in Section~\ref{s:tpu_evaluation}.

\subsection{Architecture Overview}\label{s:overall_archicteture}
E-PUR consists of four Compute Units (CUs) whose overall structure is shown in Figure~\ref{f:computational_unit}. The main components of each CU are a dot product unit (DPU) and
a Multi-functional Unit (MU). The DPU is tailored to the computations of the matrix-vector multiplications between the input sequences ($x_t$ and $h_{t-1}$) and the weights. On the other hand, the MU is used to compute activation functions (i.e., sigmoid and hyperbolic tanh) and scalar operations. Note that the computations in E-PUR are performed using either 8 or 16 bits.

In E-PUR, all gates are evaluated in parallel. For a given time-step $x_t$ of an input sequence $X$, the following steps are performed to compute the output vector $h_t$. First, 
for each output element (i.e, $n_k$) of $h_t$, the input and weight vectors are 
split into \textit{K} sub-vectors of size \textit{N}. Then, two sub-vectors of size \textit{N} are loaded from the input and weight buffers, respectively, and the dot product between them is computed by the DPU, which also accumulates the result. Next, the steps are repeated for the next $k^{th}$ 
sub-vector, and its result is added to the previously accumulated partial dot product. This process is repeated until all \textit{K} sub-vectors are computed and added together. Once the output value $n_k$ is computed, the DPU sends it to the MU where bias and scalar calculations are performed. Finally, the MU computes the activation function and stores the result in the on-chip memory for intermediate results. Finally, these steps are repeated until all the elements of $h_t$ are evaluated.

\subsection{Supporting Batching}\label{s:supporting_batching}
The base architecture of E-PUR does not support batching. Hence, in order to support the evaluation of multiple input sequences in a batch, two main modifications are done. First, E-PUR includes an on-chip memory to store intermediate results, which saves accesses to main memory and, therefore, it improves energy efficiency~\cite{fsilfa2018epur}. Regarding the intermediate results, 
most of them are generated after computing the output vector ($h_t$) since it is used as input to the next time-step of execution and as input to the next layer. However, 
including an on-chip memory for intermediate results during batching is unfeasible since the memory requirements will be huge (hundreds of megabytes). Note that the output generated on each time-step must be kept in memory since the next layer consumes it and, hence, 
intermediate results quickly become large when increasing batch size.
Hence, a vast amount of intermediate results are generated for large 
batch sizes (e.g. $time\_steps*batch\_ size*model\_neurons)$. Consequently, intermediate results among RNN layers are stored in the main memory. Despite this, RNN batching improves weight locality largely since weights are fetched once and reused to compute multiple input sequences in parallel, reducing main memory accesses for fetching weights. This trade-off is highly lucrative since the memory footprint for the weights is typically much larger than the size of the intermediate results. For instance, on E-PUR, when sequence padding is employed with a batch size of 64, energy consumption and performance are improved by 3.15x and 36x on average, respectively.

The second extension to the baseline architecture is the addition of extra DPUs and MUs on each CU, as shown in Figure~\ref{f:computational_unit}. On each CU, we include \textit{N} DPUs and MUs such that we have the necessary hardware needed to evaluate \textit{N} input sequences in parallel (for a batch of size \textit{N}). Hence, a lane in E-PUR consist of a DPU and a MU. Regarding the on-chip memories, each lane includes its private input buffer, whereas the weight buffer is shared among all the lanes.

After these modifications, the evaluation of a batch is performed as follows. First, once a batch of sequences is created by the runtime (discussed in Section~\ref{s:runtime_support}), it is sent to E-PUR. In the accelerator, each input sequence in the batch is distributed to a lane. Then, on each lane, the evaluation of a sequence is done, as explained in Section~\ref{s:overall_archicteture}. Note that since all the lanes share the weight buffer, weights are broadcast to each of them where they are multiplied by their corresponding input sequence. Finally, for each lane, once the output $h_t$ of an input sequence is computed, the result is stored in the main memory.

\section{E-Batch}\label{s:technique}

\begin{figure}[t!]
	\centering
	\includegraphics[width=3.375in]{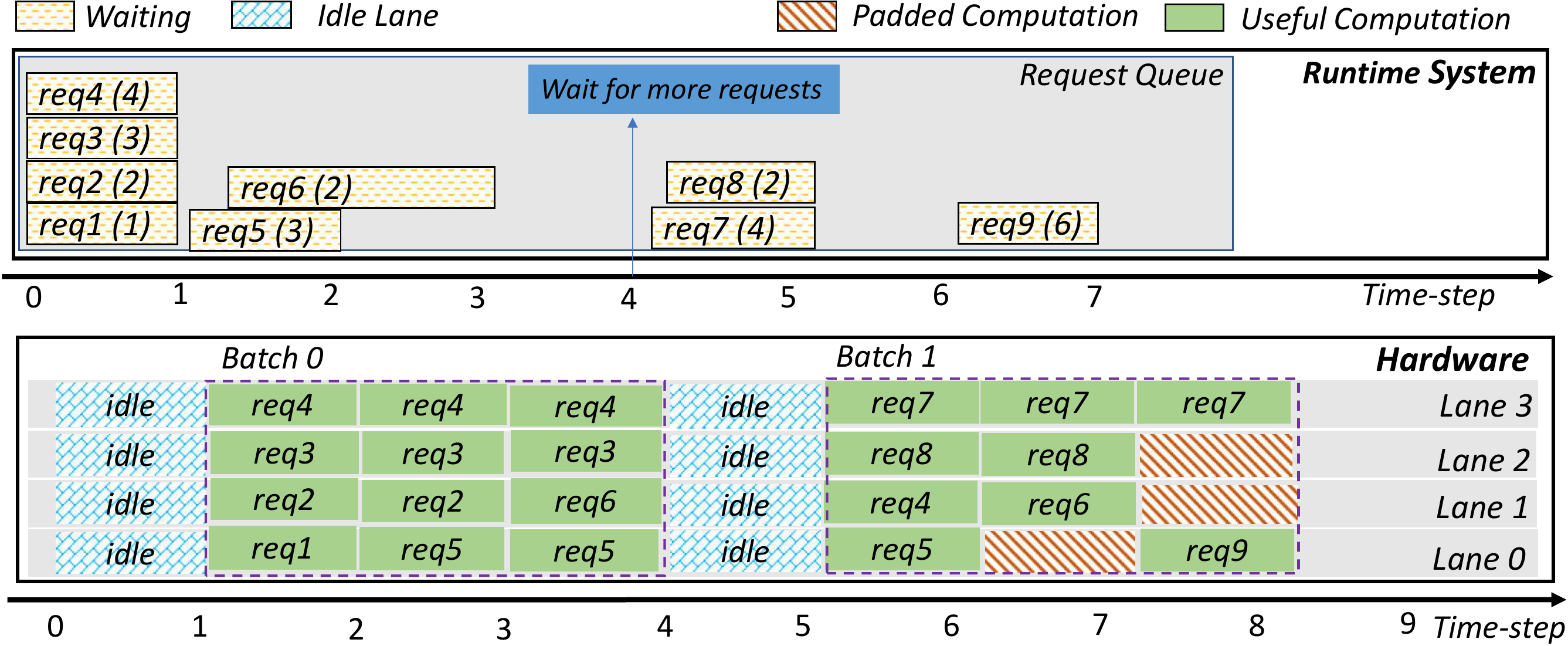}
		\vspace*{-2mm}
	\caption{At time-steps 2 and 3, \textit{lane 0} and \textit{1} become idle and the evaluation of \textit{requests 5 and 6} starts. The maximum amount of time-steps per lane ($N$) is set to 3.}
	\label{f:ebatch_example}
\end{figure}

\subsection{Overview}\label{s:technique_overview}
In section~\ref{s:batching_inefficiencies}, we described the primary sources of inefficiencies in RNN batching. In order to mitigate these issues, we propose E-Batch, a novel RNN batching scheme that tries to optimize both latency and energy consumption. In E-Batch, the creation of a batch is managed on the CPU by a runtime system, whereas the RNN computations are done in an RNN accelerator (E-PUR, TPU, ...). In contrast to other solutions, in our scheme, batches are created considering all the available requests. Since our main target is energy efficiency, E-Batch increases weight reuse by using all the available requests. When the number of requests being batched is larger than the number of available lanes, multiple requests are assigned to a given lane and are executed sequentially. In this case, the number of time-steps for a given lane is defined as the sum of the time-steps of all the requests assigned to that lane.

Since the number of time-steps for each request tends to be different, the number of time-steps on each lane also differs. To decrease the amount of padded computations (see Section~\ref{s:sequence_variability}),
our scheme tries to minimize the difference in time-steps between each lane by distributing the available requests among lanes in a way that their number of time-steps are as similar as possible. Note that this problem is an instance of the Multi-Way Number partitioning problem, which is known to be NP-Complete~\cite{korf2009}. We employ a greedy heuristic to assign each request to a given lane. For example, given the requests with the following number of time-steps (4,5,6,8,7) and a system with two lanes, we first sort all the available requests in descending order by their number of time-steps: (8,7,6,5,4). Next, we process the requests sequentially in descending order, assigning each request to the lane with the minimum amount of time-steps (e.g., 8 to the first lane, 7 to second, 6 to the second, and so on). In the aforementioned example, the first lane receives requests with time-steps (8,5,4), and the second lane gets requests with time-steps (7,6). 

Greedy partitioning does not always yield an optimal solution~\cite{korf2009}. As a result, some sequences may need to be padded. In this case, we use hardware assistance to either power gate the lanes that will perform the evaluations of the padded sequences or to assign new requests to these lanes. To this end, when a lane finishes the evaluation of all the time-steps assigned to it, E-PUR queries the runtime for a new request and, assuming that a request is available, the runtime sends it to E-PUR where its evaluation starts immediately to avoid padding. On the contrary, if there is no request available, the lane is power gated. Due to data dependencies, time-steps must be evaluated in order and cannot be distributed to multiple lanes. 

In order to reduce the impact of padding and improve throughput, E-Batch allows new requests to join a batch, but only while the first RNN layer is being processed. If a new request arrives at the server while the accelerator is processing the first (i.e., input) layer of an RNN, E-batch's runtime system tries to add the request to the current batch to reduce padded time-steps. Once the computation of the first layer for a given batch is finished, the batch is locked, and it cannot be modified for the subsequent RNN layers. The rationale behind this decision is to guarantee that all the requests in a batch belong to the same RNN layer and are processed simultaneously to maximize weight reuse, which results in significant energy savings, as shown in Section~\ref{s:results}.

In E-Batch, the number of time-steps processed by a given lane is limited to a threshold (i.e., $N$). By using a threshold, we can trade latency for energy consumption. Batching with a small number of time-steps per lane decreases latency but incurs in a large amount of weight swaps, which decreases weight locality. On the contrary, batching with a large number of time-steps per lane increases weight reuse, which significantly reduces energy consumption at the expense of an increase in latency.

When a batch is being evaluated for the first layer of the RNN, E-Batch proceeds to compute the next layer after $N$ time-steps have been evaluated per lane. If the amount of time-steps in a lane is larger than $N$, the remaining time-steps for those requests assigned to that lane are evaluated in a future batch. In other words, the evaluation of some requests is split among different batches to allow requests to progress through the next layers in the RNN and reduce latency.

When a new batch needs to be started, and the number of requests available is smaller than the batch size, the runtime waits for $T$ milli-seconds to increase hardware utilization and weight reuse (i.e., more requests are batched together) at the expense of some penalty in latency. Furthermore, if a batch is being executed and a new request arrives while the first layer of the RNN is being processed, the runtime checks for any idle lane, and if there is any, the new request is sent to the accelerator where it is assigned to that lane. Note that, to evaluate layers after the first one is computed, the previous layer's result is needed.
Therefore, even though some lanes could be idle, if a new request arrives when a batch is being evaluated for any subsequent layer, it cannot be assigned to any of those idle lanes. Also, only requests for the same RNN model are batched together.

\begin{figure}[t!]
	\centering
	\includegraphics[width=3.375in]{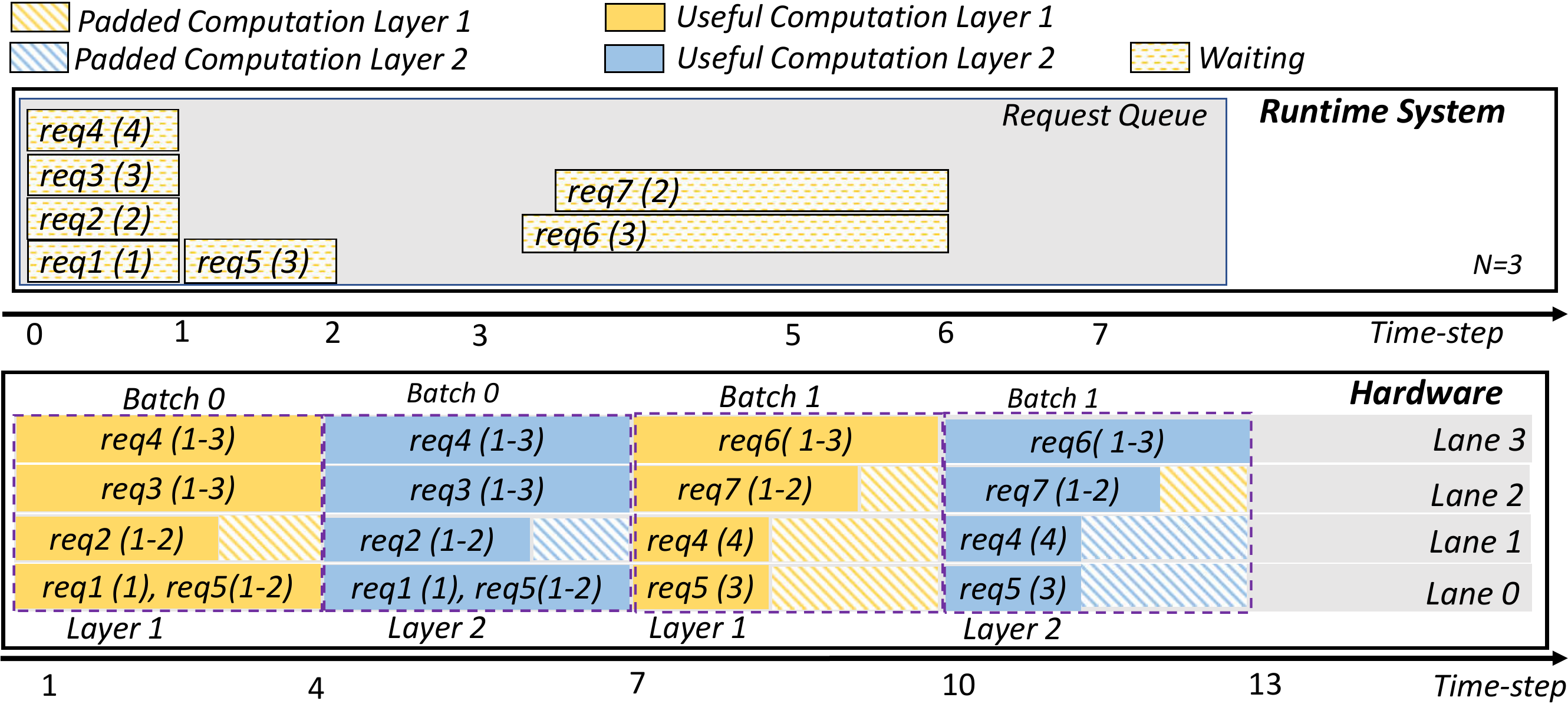}
		\vspace*{-2mm}
	\caption{Evaluation of an RNN with two LSTM layers using E-Batch. Batch 0 is evaluated for the first and second layer, before a new batch is created. \textit{req 5} arrives before finishing the evaluation of the first layer and, hence, it joins Batch 0 immediately. On the contrary, \textit{req 6-7} arrives after the evaluation of the first layer, so its evaluation is deferred until a new batch is created. }
	\label{f:ebatch_example2}
\end{figure}

 Figure~\ref{f:ebatch_example} illustrates the execution flow of E-Batch. Like the previous examples, a batch size of 4 is used and an LSTM model with one layer. Also, a value of 3 time-steps is used for $N$, whereas $T$ is set to two time-steps. At the beginning, \textit{requests 1-4} are in the waiting queue. Since the number of requests in the queue is four, a new batch (\textit{batch 0}) is created by the runtime, assigning each request in the batch to a given lane using Greedy partitioning. Afterwards, \textit{batch 0} is sent to the accelerator in time-step 1. At time-step 2, the execution of \textit{request 1} is completed. Hence, the accelerator sends a signal to the runtime to indicate that \textit{lane 0} is idle. Thus, \textit{request 5} is assigned to \textit{lane 0} since it is the oldest request in the waiting queue. Similarly, \textit{request 6} is assigned to \textit{lane 1} after \textit{request 2} is finished. At time-step 4, the execution of \textit{batch 0} is completed since $N$ is 3. At this point, \textit{requests 1-2} are completed, and the RNN output is sent back to the client. However, for \textit{requests 4-6}, only 3, 2, and 1 time-steps have been evaluated. Therefore, the remaining time-steps will be computed in a future batch. After the evaluation of \textit{batch 0}, only \textit{requests 4-6} are in the queue. Hence, since the number of requests on the queue is smaller than the number of lanes, the runtime waits for $T$ time-steps or enough requests to fill all the lanes.

 As seen in Figure~\ref{f:ebatch_example}, at time-step 5 \textit{request 4-8} are in queue, hence \textit{batch 1} is created by the runtime. In this case, since the number of requests in the queue is greater than the number of lanes, \textit{requests 4 and 6} are assigned to lane 1. Finally, \textit{batch 1} is sent to the accelerator, where its evaluation is performed in manner similar to \textit{batch 0}.
 
 Figure~\ref{f:ebatch_example2} shows the behavior of our scheme for RNN models with more than one layer. Similar to the previous example, for time-step 1, \textit{requests 1-4} are batched and sent to the accelerator. Furthermore, at time-step 2, \textit{request 1} is completed and \textit{request 5} is assigned to \textit{lane 0}. Since $N$ is 3, after three time-steps have been evaluated, we proceed with the evaluation of the next layer. Note that \textit{requests 6-7} are available at time-step four and that lane 1 is idle after time-step 6. However, as mentioned before, new requests cannot join a batch after the first layer's evaluation is completed. Hence, \textit{requests 6-7} wait until the evaluation of \textit{batch 0} finishes for the second layer. After \textit{batch 0} is completed, the output results for \textit{requests 1-2} are sent to the user, whereas a new batch is created using the remaining time-steps of \textit{request 4-5} and the new requests \textit{request 6-7}.

\begin{figure}[t!]
	\centering
	\includegraphics[width=3.375in]{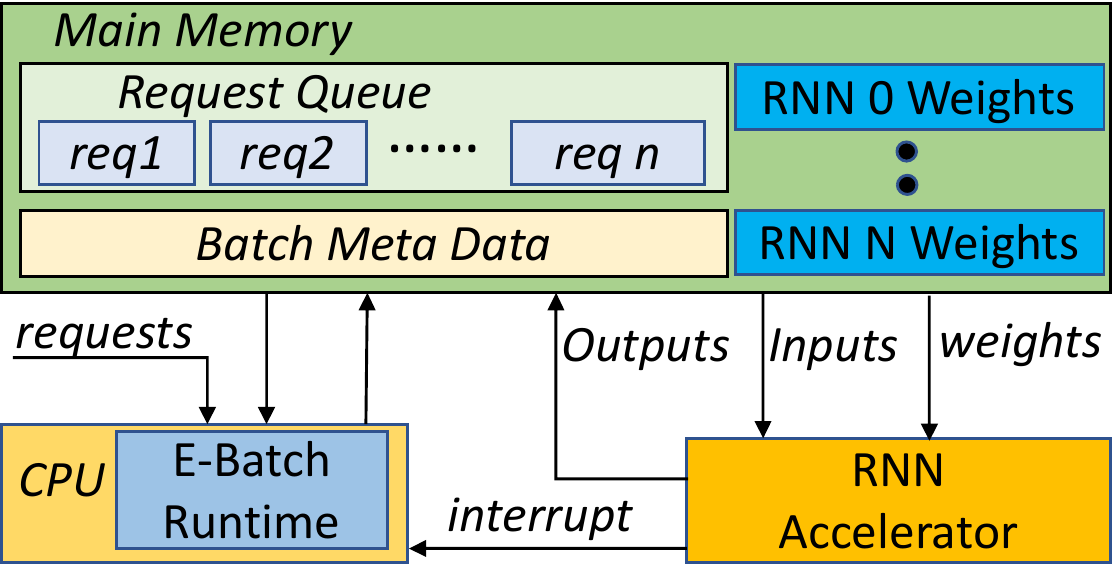}
		\vspace*{-3mm}
	\caption{Overview of E-Batch System Architecture.}
	\label{f:ebatch_arch}
\end{figure}

The overall architecture of E-Batch is shown in  Figure~\ref{f:ebatch_arch}. It is composed of a runtime system and an RNN accelerator. The runtime is in charge of creating and managing batches of requests, whereas evaluations are performed in the accelerator. In the next sub-sections, we describe in more detail these components.

\subsection{Hardware support}\label{s:hardware_support}

In order to support E-Batch in E-PUR, the following modifications are done.
First, we include an interrupt, which is used to signal the runtime when a lane becomes idle. Furthermore, we include a small buffer (i.e., \textit{the request buffer}) to keep track of the lane where each request is being evaluated. Also, in \textit{the request buffer}, we store the number of time-steps that have to be processed for each request. Finally, we include a register to store the parameter $N$ (i.e., maximum number of time-steps).
Once a batch is received, the evaluation of the first layer works as follows. For each request in the batch, an entry in the \textit{request buffer} is created and initialized with its meta-information. Then, when a time-step is evaluated, the number of time-steps processed for the active requests are updated in the \textit{request buffer}. If all the time-steps for a given request are completed, we proceed with the next request in the lane. However, if there are no more requests in the lane, a signal is sent to runtime to indicate that a lane is idle so that a new request can join the current batch being executed.  After $N$ time-steps have been evaluated, we continue with the computation of the next layers. Finally, once the evaluation of a batch is completed, the number of time-steps evaluated for each request is sent back to the runtime. 

\subsection{Runtime support}\label{s:runtime_support}
The runtime is in charge of the management and creation of batches. It includes a queue where new requests
arriving at the system are stored. Furthermore, for each request, the runtime tracks the number of time-steps that have been evaluated. Also, it knows whether the accelerator is processing or idle.

When the accelerator is idle, the system will create a new batch using all the available requests and employing the Greedy
partitioning algorithm described in~\ref{s:technique_overview}. Then, once a batch is created, for each batched request, the lane assigned to it and the number of time-steps that need to be evaluated are sent to the accelerator. Once the evaluation of a batch is completed, the number of time-steps evaluated are updated for all its requests. Moreover, requests that are completed (i.e., all its time-steps have been executed) are returned to the user. On the other hand, new requests and requests with time-steps pending for evaluation are batched together. If the available requests are not sufficient to create a complete batch, the runtime will wait until the batch is completed or $T$ milli-seconds have passed.

 \begin{table*}[t!]
	\caption{RNN Networks used for the experiments.}
	\label{t:lstm_networks}
	\centering
	\begin{tabular}{cccccc}
		\cellcolor[gray]{0.9}\small\textbf{Network}&\cellcolor[gray]{0.9}\small\textbf{App Domain}&\cellcolor[gray]{0.9}\small\textbf{Cell Type}&\cellcolor[gray]{0.9}\small\textbf{Layers}&\cellcolor[gray]{0.9}\small\textbf{Cell Size}&\cellcolor[gray]{0.9}\small\textbf{ Dataset }  \\
		\small DeepSpeech2~\cite{deepspeech2}&\small Speech Recognition&\small GRU&\small 5&\small 800&\small LibriSpeech\\
		
		\cellcolor[gray]{0.9}\small MNMT~\cite{britzGLL17}&\cellcolor[gray]{0.9}\small Machine Translation&\cellcolor[gray]{0.9}\small LSTM&\cellcolor[gray]{0.9}\small 8&\cellcolor[gray]{0.9}\small 1024&\cellcolor[gray]{0.9}\small WMT'15 En $\rightarrow$ Ge \\
	\end{tabular}
\end{table*}

\begin{table}[t!]
	\caption{Hardware configuration.}
	\label{t:hardware_params}
	\centering
	\begin{tabular}{cc}
		\hline
		\multicolumn{2}{c}{\textbf{E-PUR}}\\
		\cellcolor[gray]{0.9}\small\textbf{Parameter}&\cellcolor[gray]{0.9}\small\textbf{Value}\\
		\small Technology&\small 28 nm\\
		\cellcolor[gray]{0.9}\small Frequency&\cellcolor[gray]{0.9}\small 500 MHz\\
		\small Weight Buffer&\small 2 MiB per CU\\
		\cellcolor[gray]{0.9}\small Input Buffer&\cellcolor[gray]{0.9}\small 128 KiB per CU\\
	    \small DPU Width&\small 64 operations\\
	    	\hline
		\multicolumn{2}{c}{\cellcolor[gray]{0.9}\textbf{TPU-like accelerator}}\\
		
	    \small Frequency&\small 700 MHz\\
		\cellcolor[gray]{0.9}\small SRAM  Buffer&\cellcolor[gray]{0.9}\small 24 MiB \\
	     \small Systolic Array PEs&\small 128x128 \\

	\end{tabular}
\end{table}

\subsection{Supporting E-Batch On a TPU-Like Architecture}

TPU is a state-of-the-art accelerator for neural networks~\cite{jouppi2017TPU}. It is composed of a systolic array of processing elements (PEs) and
on-chip memory for weights and activations. In order to evaluate LSTM models on TPU like architectures, an output stationary dataflow 
is employed~\cite{rieraCGPA2019,samajdar2018scale}. In this regard, for an RNN model with $N$ neurons, neuron $n_k$ will be mapped to all the PEs in column $k$ of the array. Regarding the weights, they are mapped to the columns of the systolic array, whereas the input sequences are mapped to
rows of the array (one input sequence per row). Thus, to evaluate $n_k$, its weights are streamed down through column $k$, whereas the elements of each input sequence are streamed through the rows. Note that the maximum batch size supported is equal to the number of rows in the array and that each of them will correspond to a \textit{processing lane}. To support E-Batch in a TPU-like 
 architecture, as done in E-PUR, we add an interrupt to signal the end of a sequence to the runtime and a \textit{request buffer} as described in~\ref{s:hardware_support}.

\section{Evaluation Methodology}\label{s:methodology}

To evaluate our proposal, we use the state-of-the-art LSTM networks shown in Table~\ref{t:lstm_networks}.  
They consists of RNNs for popular applications such as machine translation~\cite{wu2016google} and 
speech recognition~\cite{deepspeech2}. The input sequences for these RNNs have sizes varying from ten to a few thousand time-steps. Also, they have a different number of layers and the hidden dimension.

In order to estimate performance and energy, we use a cycle-level simulator of E-PUR and TPU, using the configuration parameters in Table~\ref{t:hardware_params}. We model a TPU-like architecture using the SCALE-Sim~\cite{samajdar2018scale} simulator. We set the number of filters as the number of neurons in the model and the input features as the number of weights per neuron. The width and height of the filters are set to one, whereas the width of the input features is set to the batch size.

To estimate energy consumption, we modeled the different components of E-PUR and TPU in Verilog and synthesized them using Synopsis Design Compiler and to determine the static and dynamic energy consumption of on-chip memories we employed CACTI~\cite{muralimanohar2009cacti}. Finally, for main memory, we used the MICRON power model~\cite{micron} and modeled a 4GB LPDDR4 DRAM.

Regarding the runtime, we implemented a system that employs the timing estimation and status of the accelerator' simulator (E-PUR or TPU) to batch new requests. For our experiments, we simulate several hours of the system execution. To this end, the train/test datasets are not sufficient since they would be processed in a short time span. Aiming to generate more requests to be processed by our system, we analyzed the distribution of the number of time-steps for each input sequence in the train and test set of the RNN models in the benchmarks. Based on this analysis, we generate requests whose number of time-steps follows the distribution observed in the original datasets. Note that the execution and energy consumption of a given request only depends on the RNN parameters and the number of time-steps. 
To reproduce a real environment, we simulate the arrival time of each user's request. To this end, we use a Poisson distribution. In order to increase or decrease the number of requests per second arriving to the system (i.e., system load), we change the average inter-arrival time between requests. Note that, for each of our experiments, the average request arrival rate is kept constant.

\section{Experimental Results}\label{s:results}

This section presents the evaluation of our proposal on an E-PUR like accelerator and a TPU-like architecture.
Regarding the batch size, we tested several batch sizes and found that the results are similar. 
Thus, we chose 64 as it delivered the best trade-off between performance and area for E-PUR, 
whereas for a TPU-like architecture, we use 128.  
The rest of this section is organized as follows. First, we discuss the evaluation of bucketing and
cellular batching on E-PUR. Second, we discuss the performance and energy consumption of E-Batch for E-PUR. 
Finally, we discuss the results of E-Batch for a TPU-like architecture.

\subsection{Sequence bucketing and cellular batching on an E-PUR-like Accelerator}\label{s:cellular_an_bucketing_results}

Regarding Bucketing, our experiments show that sequence padding is, on average, 7x faster than bucketing and delivers 1.6x higher requests per joule for the machine translation model in Table~\ref{t:lstm_networks}. As discussed in Section~\ref{s:resource_utilization}, bucketing with a small \textit{bucket width} 
  tends to create batches with a low number of requests and, thus, weights are swapped more frequently, resulting in larger energy consumption than sequence padding, which manages to create batches with a larger number of requests.
   Moreover, with bucketing, requests have to wait longer before they are sent to the accelerator, thus increasing their latency.
   Note that when the number of requests per second that arrives at the system is low, both bucketing and padding achieve similar results since they behave similarly.
   
   Regarding cellular batching, it achieves a maximum throughput of 100 requests per second for the GNMT model, shown in Table~\ref{t:lstm_networks}. 
Moreover, it consumes 20x more energy per request than sequence padding. Also, on average, the latency is 900ms.
As mentioned in Section~\ref{s:weight_locality}, cellular batching incurs in a large amount of weight swapping when evaluating 
RNN models with more than one layer. As previously shown in Figures~\ref{f:energy_cell_pad_buck} and~\ref{f:bytes_read_cell_pad}, as the number of layers increases
the number of bytes read per request increases and thus, energy efficiency decreases.

\subsection{E-Batch on an E-PUR-like Accelerator}\label{s:epur_evalaution}

\begin{figure}[t!]
	\centering
	\includegraphics[width=3.375in]{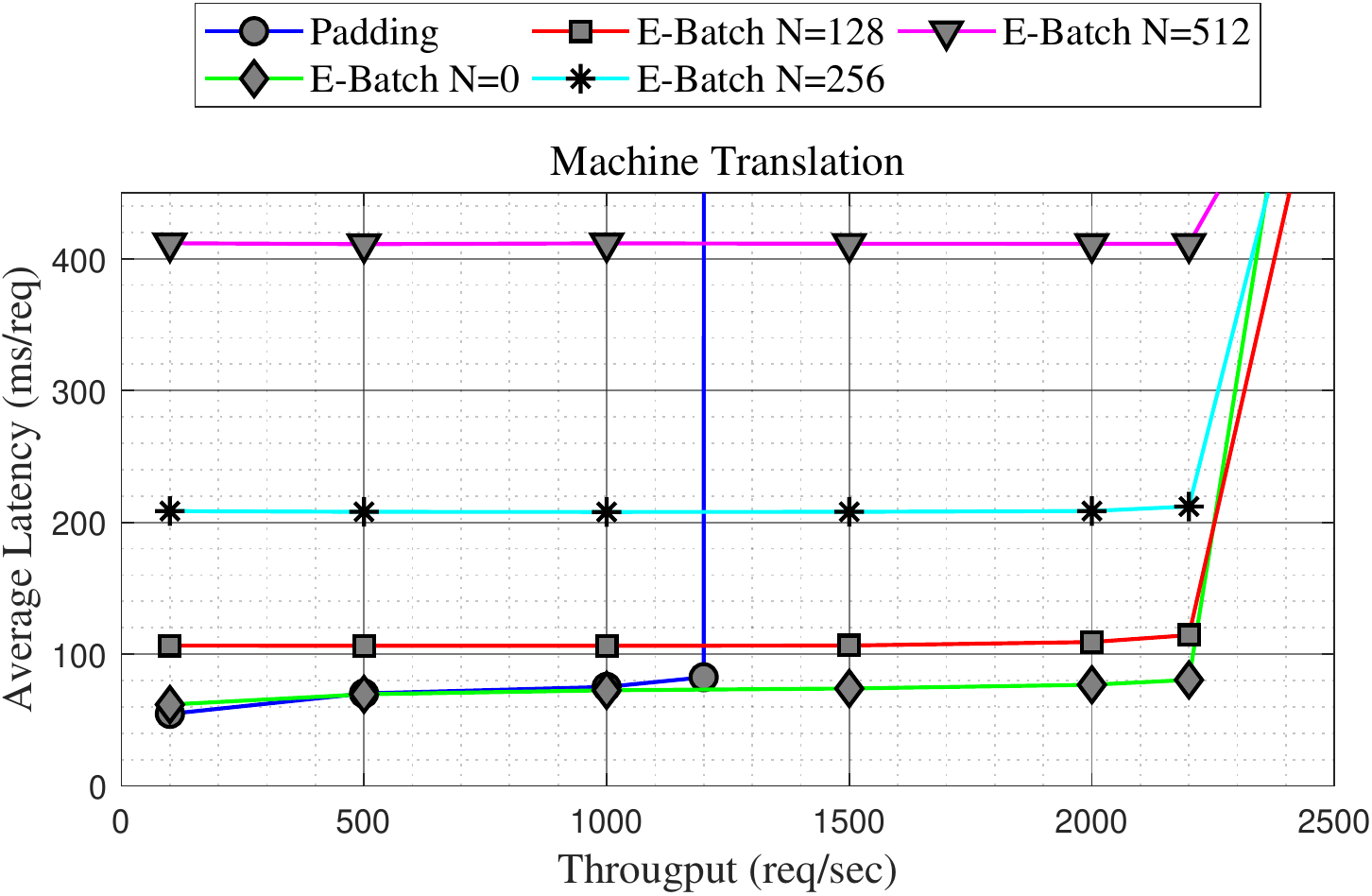}
		\vspace*{-3mm}
	\caption{Average Latency vs Throughput for Machine Translation~\cite{britzGLL17} using a batch size of 64 on E-PUR.}
	\label{f:epur_nmt_latency}
\end{figure}

\begin{figure}[t!]
	\centering
	\includegraphics[width=3.375in]{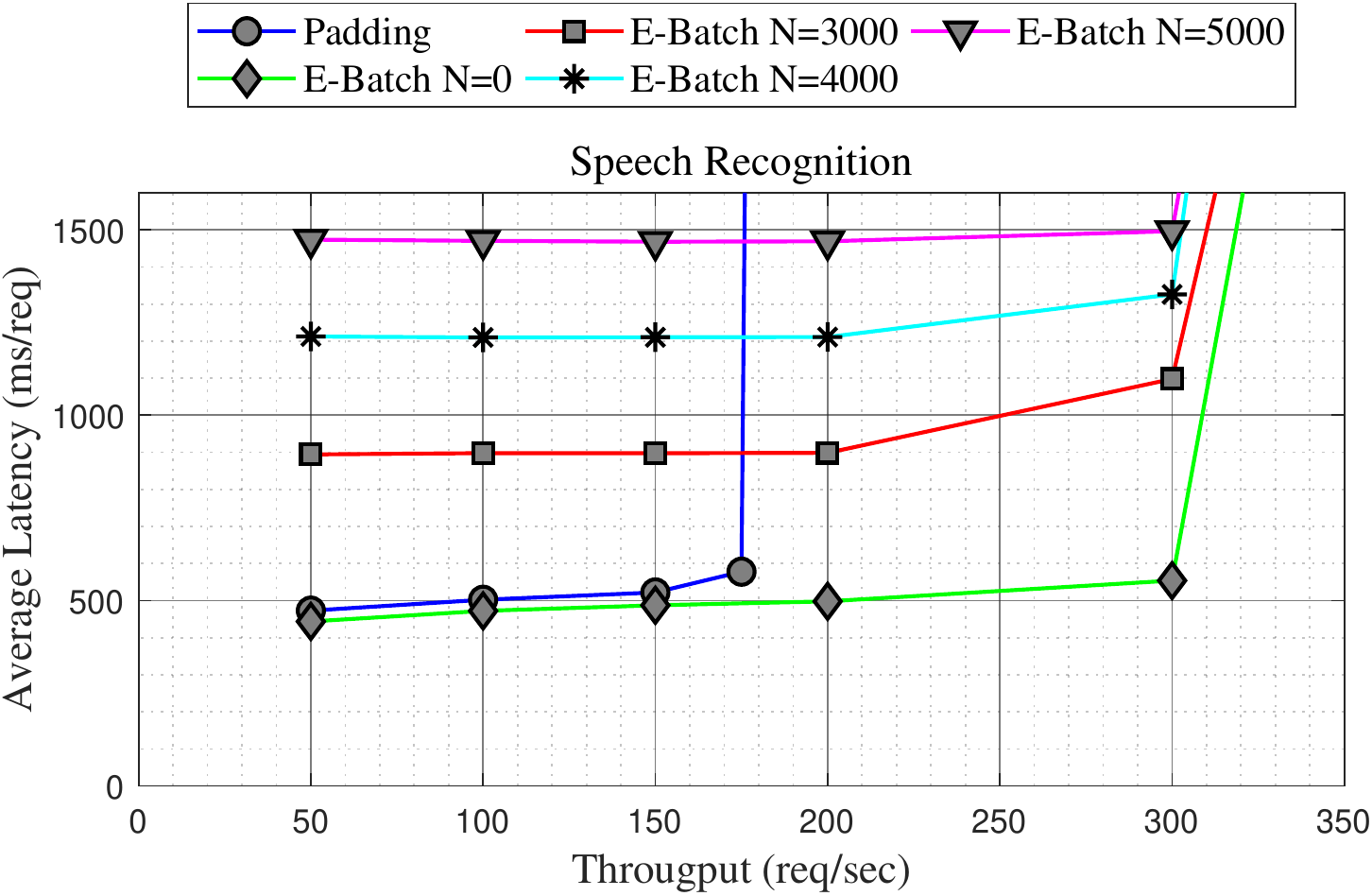}
		\vspace*{-3mm}
	\caption{Average Latency vs Throughput for Speech Recognition~\cite{deepspeech2} using a batch size of 64 on E-PUR.}
	\label{f:epur_deep_latency}
\end{figure}

Figures~\ref{f:epur_nmt_latency} and \ref{f:epur_deep_latency} show the 
average latency per request for Machine Translation and Speech Recognition RNNs, respectively, on an E-PUR-like accelerator. 
Both figures include the baseline and our E-batch scheme under different loads and using 64 lanes (maximum batch size). For E-Batch, we use different values of the threshold $N$ (i.e., maximum number of time-steps in a lane) to evaluate the trade-off between energy and latency. In these plots and others, $N=0$ refers to the E-Batch configuration where the maximum number of time-steps in a lane is set to the number of time-steps of the longest sequence in the batch when it is created.
The Speech Recognition network has a larger average latency since its input sequences are typically larger than the input sequences of Machine Translation model; therefore, they result in more considerable processing and memory transfer times.

As can be seen for both networks, when $N$ is zero, the latency of using sequence padding and E-Batch are similar because, in this case, the longest sequence in a given batch is the same for both schemes. However, since E-Batch can add new requests to a batch while it is being processed, the queuing time decreases, henceforth reducing the average latency. Note that for low loads (i.e., 100 requests per second in Figure~\ref{f:epur_nmt_latency}), the average latency is sightly higher for E-Batch.  The reason is that E-batch waits for some time to increase the number of requests that are batched together, thus improving weight locality and energy efficiency.

\begin{figure}[t!]
	\centering
	\includegraphics[width=3.375in]{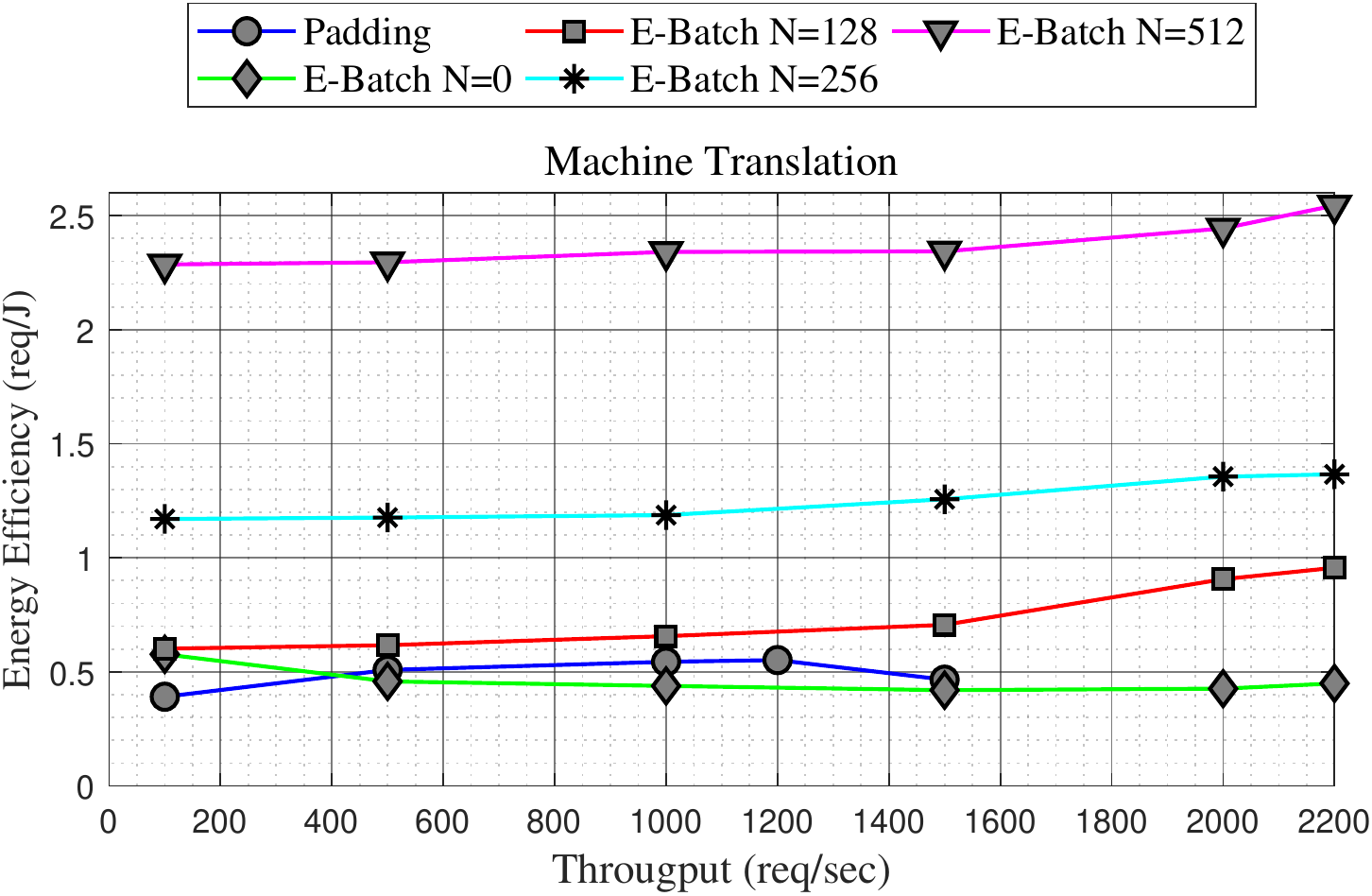}
		\vspace*{-3mm}
	\caption{Average number of Requests per Joule vs Throughput for Machine Translation~\cite{britzGLL17} using a batch size of 64 on E-PUR.}
	\label{f:epur_nmt_req_joule}
\end{figure}

Regarding the maximum throughput, as it can be seen in Figures~\ref{f:epur_nmt_latency} and \ref{f:epur_deep_latency} E-Batch achieves 1.83x and 1.77x improvement over padding for Machine Translation and Speech Recognition, respectively. This improvement in throughput comes from allowing new requests to start execution while the first layer of an RNN is being evaluated. Also, because of the variability in the number of time-steps, when the evaluation of small requests is completed, a new request from the waiting queue can join the current batch being executed. Therefore, hardware utilization is increased, improving system throughput. The maximum throughput obtained by E-Batch for different values of $N$ in Figures~\ref{f:epur_nmt_latency} and \ref{f:epur_deep_latency} are similar, slightly increasing when $N$ becomes very large.

\begin{figure}[t!]
	\centering
	\includegraphics[width=3.375in]{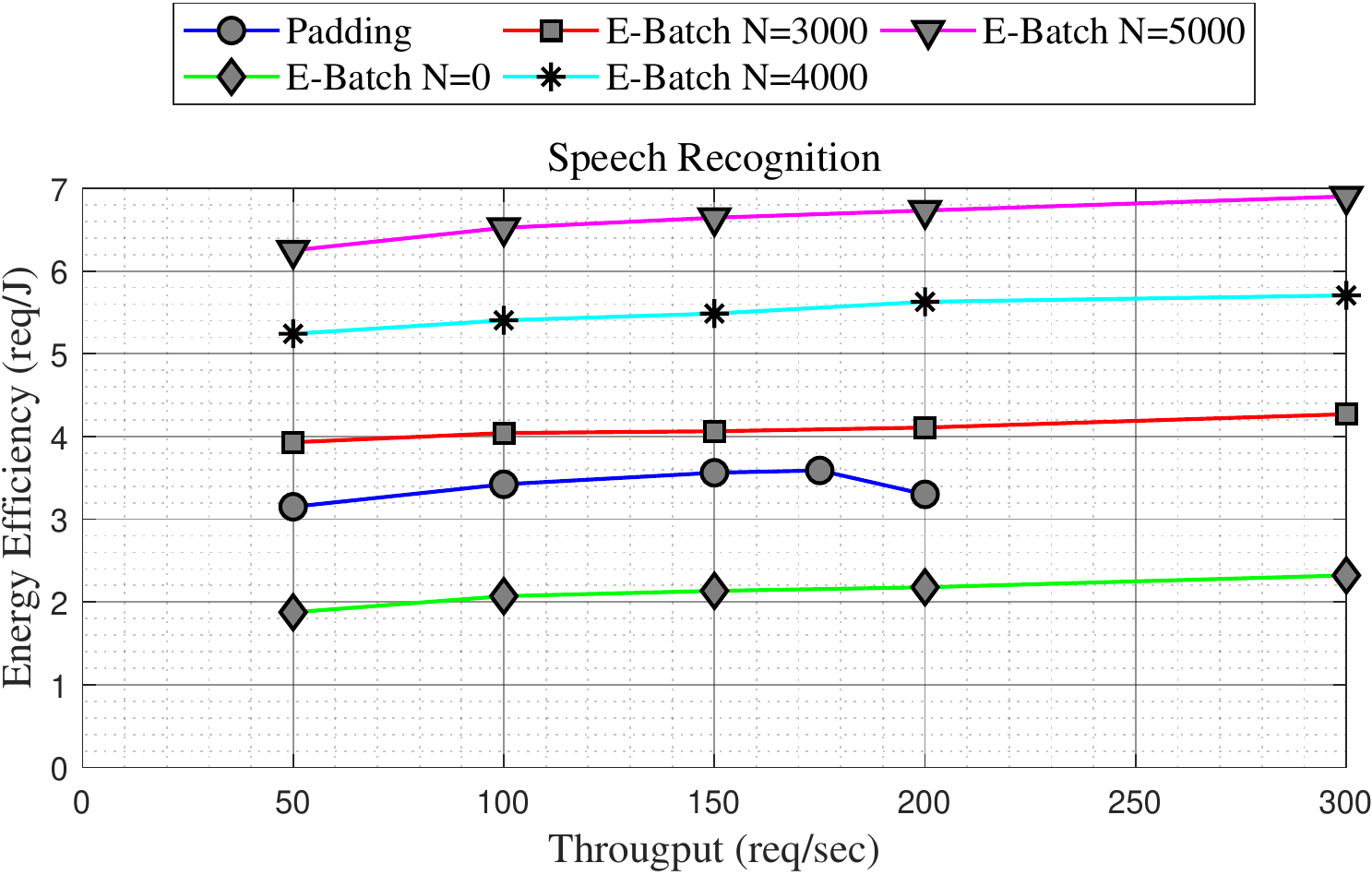}
		\vspace*{-3mm}
	\caption{Average number of Requests per Joule vs Throughput for Speech Recognition~\cite{britzGLL17} using a batch size of 64 on E-PUR.}
	\label{f:epur_deep_req_joule}
\end{figure}

Figures~\ref{f:epur_nmt_req_joule} and \ref{f:epur_deep_req_joule} show the average number of requests per joule for Machine Translation and Speech Recognition, respectively. As can bee seen, when $N$ is zero, sequence padding tends to have a slightly better energy efficiency than E-Batch for low loads. The reason is that when the number of requests in a batch is small, as new requests arrive at the system, they are added to the batch being processed. However, since $N$ is 0 once the time-steps for the longest and oldest request in the batch are computed, the system proceeds to evaluate deeper layers. As a result, the computations of some requests are divided among several batches, and they require more memory accesses to fetch the weights. On the contrary, as the load increases, batches with a larger amount of time-steps are created, and thus, weight reuse increases.

As shown in Figures~\ref{f:epur_nmt_req_joule} and \ref{f:epur_deep_req_joule}, increasing the value of $N$ will 
dramatically increase the energy efficiency of E-Batch for both networks. The reason for this is that
since the system waits until $N$ time-steps are evaluated in the first layer, a large amount of time-steps 
are batched together and, as a result, weight reuse is increased.

\subsection{E-Batch on a TPU-like Accelerator}\label{s:tpu_evaluation}

\begin{figure}[t!]
	\centering
	\includegraphics[width=3.375in]{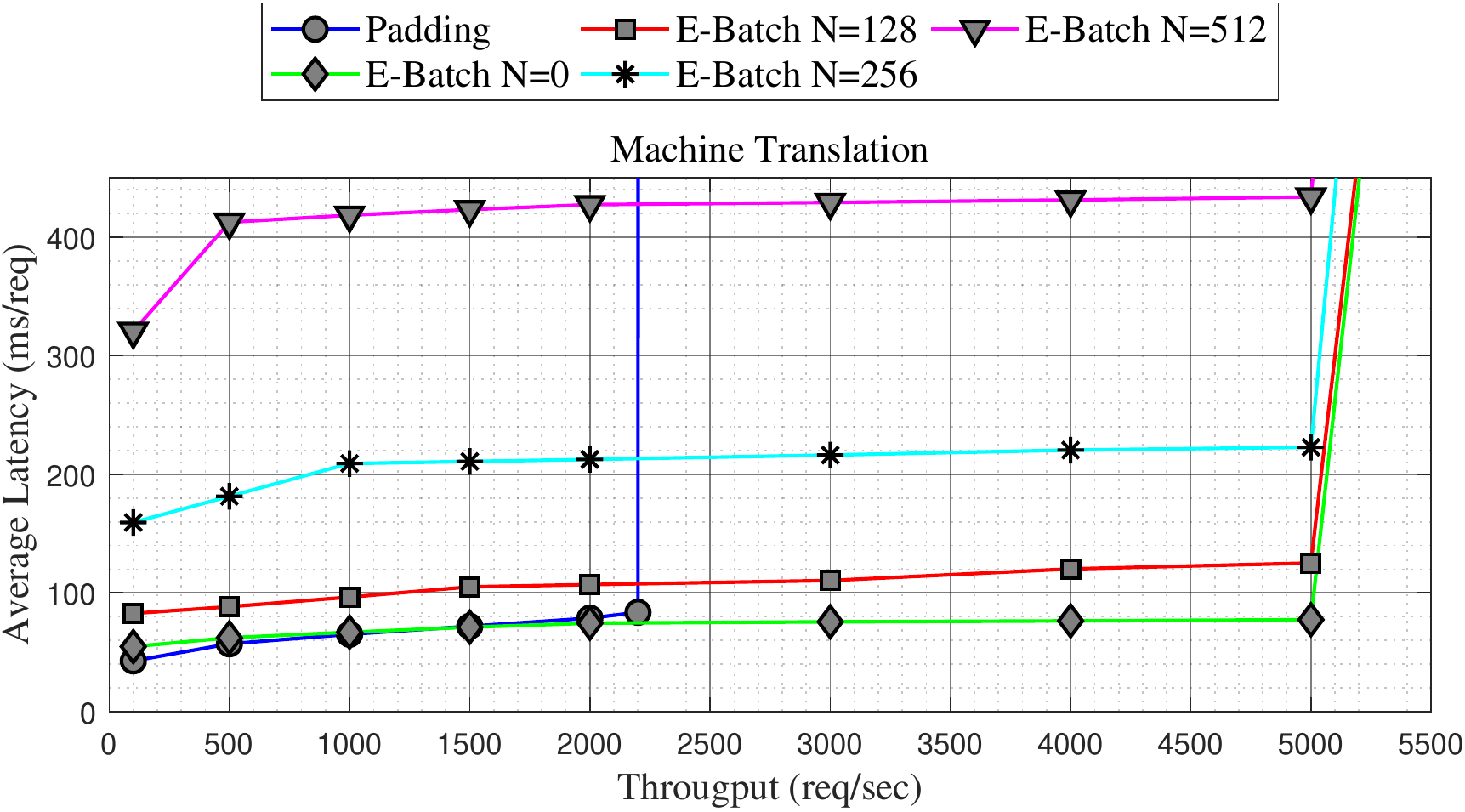}
		\vspace*{-3mm}
	\caption{Average Latency vs Throughput for Machine Translation~\cite{britzGLL17} using a batch size of 128 for TPU.}
	\label{f:tpu_nmt_latency}
\end{figure}

Figure~\ref{f:tpu_nmt_latency} shows the average latency per request for sequence padding and E-Batch running on top of a TPU-like accelerator. The number of lanes used is 128 since this is the number of rows of the systolic array in the TPU. As can be seen, when $N$ is zero, the average latency of sequence padding and E-Batch are similar since, in this case, the largest number of time-steps for a given batch is the same in both schemes.  
However, since E-Batch allows new requests to join a batch while it is being computed for the first layer, the maximum throughput of E-Batch is 2.1x higher.

\begin{figure}[t!]
	\centering
	\includegraphics[width=3.375in]{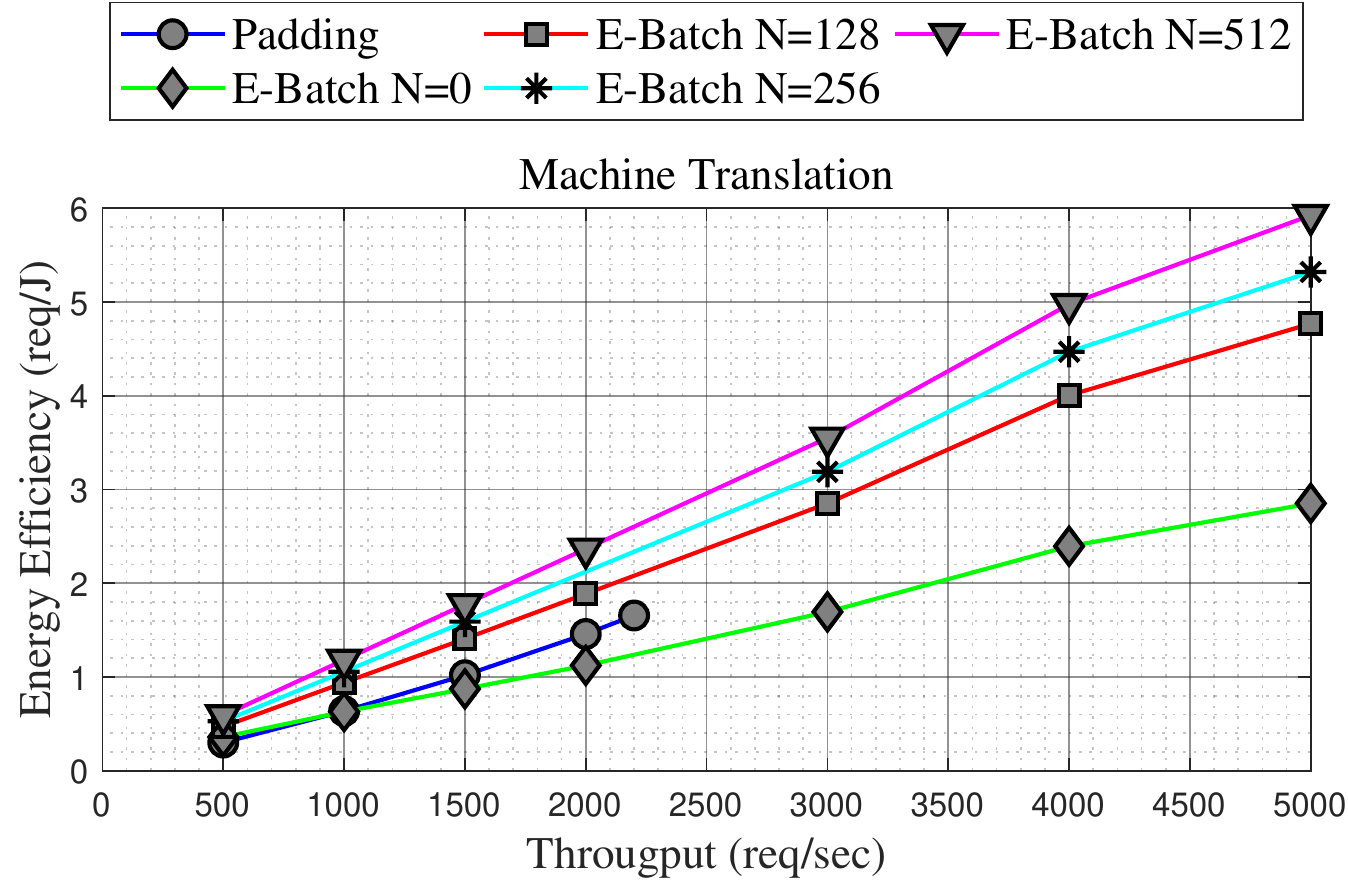}
	\vspace*{-3mm}
	\caption{Average number of Requests per Joule vs Throughput for Machine Translation~\cite{britzGLL17} using a batch size of 128 for TPU.}
	\label{f:tpu_nmt_req_joule}
\end{figure}

Figure~\ref{f:tpu_nmt_req_joule} shows the number of requests per joule for the machine translation network evaluated on a TPU-like architecture.  Similar to E-PUR, when the load is small, and $N$ is zero, many batches with a small number of time-steps are created. As a result, the number of memory accesses increases. However, as the number of requests per second increases, a more significant number of requests are batched together, thus increasing weight reuse.
Similarly, as $N$ increases, a larger amount of time-steps are batched together, increasing weight reuse. For instance, when $N$ is 128, and the load is 2000 req/sec, E-batch achieves an energy efficiency improvement of 1.3x compared to the baseline, whereas when $N$ is 256 or 512 energy efficiency is improved by 1.46x and 1.6x, respectively. However, note that when improving energy efficiency, the average latency is also increased.

\section{Related Work}\label{s:related_work}

Recurrent Neural Networks (RNNs) have achieved tremendous success in sequence to sequence problems. Not surprisingly,
several hardware accelerators~\cite{jouppi2017TPU, brainwave2018, yazdani2019lstmsharp,  guan2017fpga } and software libraries~\cite{cudnn, deepcpu, grnn2019} tailored to improve energy efficiency and performance of RNNs have been proposed recently. In addition to hardware acceleration, several techniques such as pruning~\cite{rieraCGPA2019, HanESE2017:}, model compression~\cite{masr2019}, and computations reuse\cite{riera2018computation, fuzzy2019} have been employed. Our optimizations are orthogonal to those techniques.

Regarding software libraries such as~\cite{grnn2019, cudnn, paszke2017automatic}, they are mainly tailored to GPUs and CPUs~\cite{deepcpu}, whereas our proposal targets specialized accelerators. In order to handle sequences of differences sizes, previous proposals employ sequence padding or bucketing. In CNTK~\cite{seideCntk2016}, wasted computations are avoided by trying to batch small sequences when they can fit in the padded space. However, this is not always possible since small sequences may not be available. Conversely, in E-Batch, a sequence can be split among different batches so that only the amount of time-steps required to fill the padded space is used. 

Batching RNN sequences are supported in hardware accelerators such as BrainWave and LSTM-Sharp. BrainWave~\cite{brainwave2018} is highly optimized for batches of size one, and inputs are processed sequentially, by computing one single input at a time. In this accelerator, sequence padding is not required. However, employing batch sizes larger than two is unfeasible since inputs are processed sequentially. LSTM-Sharp~\cite{yazdani2019lstmsharp} focuses on increasing resource utilization. It addresses the issue of extra padded computations that occur when performing matrix-vector multiplications, and the number of multipliers per tile are not multiples of the input vector size. This padding is different to padding several sequences to make their sizes homogeneous. Sequence padding is also needed for LSTM-Sharp, to support batching. E-batch is orthogonal to this work and can be implemented on top of this accelerator. TPU~\cite{jouppi2017TPU} supports large batch sizes employing sequence padding.

\section{Conclusions}\label{s:conclusions}

In this paper, we presented E-Batch, a batching system for recurrent neural networks that increases throughput while also improving energy efficiency. E-Batch consists of a runtime and minor extensions to the hardware accelerator. In E-Batch, larger sequences are batched together to decrease memory accesses. Furthermore, throughput is increased by allowing requests to join other requests while their execution is ongoing. We evaluated E-Batch on top of E-PUR and TPU, two state-of-art hardware accelerators for RNNs. Our experimental results show that in E-PUR, E-Batch improves throughput by 1.8x and energy efficiency by 3.6x, whereas for TPU throughput is enhanced by 2.1x and energy efficiency by 1.6x.

\ifCLASSOPTIONcaptionsoff
  \newpage
\fi

\bibliographystyle{IEEEtran}
\bibliography{references}

\begin{thebibliography}{10}
\providecommand{\url}[1]{#1}
\csname url@samestyle\endcsname
\providecommand{\newblock}{\relax}
\providecommand{\bibinfo}[2]{#2}
\providecommand{\BIBentrySTDinterwordspacing}{\spaceskip=0pt\relax}
\providecommand{\BIBentryALTinterwordstretchfactor}{4}
\providecommand{\BIBentryALTinterwordspacing}{\spaceskip=\fontdimen2\font plus
\BIBentryALTinterwordstretchfactor\fontdimen3\font minus
  \fontdimen4\font\relax}
\providecommand{\BIBforeignlanguage}[2]{{%
\expandafter\ifx\csname l@#1\endcsname\relax
\typeout{** WARNING: IEEEtran.bst: No hyphenation pattern has been}%
\typeout{** loaded for the language `#1'. Using the pattern for}%
\typeout{** the default language instead.}%
\else
\language=\csname l@#1\endcsname
\fi
#2}}
\providecommand{\BIBdecl}{\relax}
\BIBdecl

\bibitem{britzGLL17}
\BIBentryALTinterwordspacing
D.~Britz, A.~Goldie, M.~Luong, and Q.~V. Le, ``Massive exploration of neural
  machine translation architectures,'' \emph{CoRR}, vol. abs/1703.03906, 2017.
  [Online]. Available: \url{http://arxiv.org/abs/1703.03906}
\BIBentrySTDinterwordspacing

\bibitem{deepspeech2}
\BIBentryALTinterwordspacing
D.~Amodei, R.~Anubhai, E.~Battenberg, C.~Case, J.~Casper, B.~Catanzaro,
  J.~Chen, M.~Chrzanowski, A.~Coates, G.~Diamos, E.~Elsen, J.~Engel, L.~Fan,
  C.~Fougner, T.~Han, A.~Y. Hannun, B.~Jun, P.~LeGresley, L.~Lin, S.~Narang,
  A.~Y. Ng, S.~Ozair, R.~Prenger, J.~Raiman, S.~Satheesh, D.~Seetapun,
  S.~Sengupta, Y.~Wang, Z.~Wang, C.~Wang, B.~Xiao, D.~Yogatama, J.~Zhan, and
  Z.~Zhu, ``Deep speech 2: End-to-end speech recognition in english and
  mandarin,'' \emph{CoRR}, vol. abs/1512.02595, 2015. [Online]. Available:
  \url{http://arxiv.org/abs/1512.02595}
\BIBentrySTDinterwordspacing

\bibitem{fsilfa2018epur}
\BIBentryALTinterwordspacing
F.~Silfa, G.~Dot, J.-M. Arnau, and A.~Gonz\'{a}lez, ``E-pur: An
  energy-efficient processing unit for recurrent neural networks,'' in
  \emph{Proceedings of the 27th International Conference on Parallel
  Architectures and Compilation Techniques}, ser. PACT '18.\hskip 1em plus
  0.5em minus 0.4em\relax New York, NY, USA: ACM, 2018, pp. 18:1--18:12.
  [Online]. Available: \url{http://doi.acm.org/10.1145/3243176.3243184}
\BIBentrySTDinterwordspacing

\bibitem{jouppi2017TPU}
\BIBentryALTinterwordspacing
N.~P. Jouppi, C.~Young, N.~Patil, D.~Patterson, G.~Agrawal, R.~Bajwa, S.~Bates,
  S.~Bhatia, N.~Boden, A.~Borchers, R.~Boyle, P.-l. Cantin, C.~Chao, C.~Clark,
  J.~Coriell, M.~Daley, M.~Dau, J.~Dean, B.~Gelb, T.~V. Ghaemmaghami,
  R.~Gottipati, W.~Gulland, R.~Hagmann, C.~R. Ho, D.~Hogberg, J.~Hu, R.~Hundt,
  D.~Hurt, J.~Ibarz, A.~Jaffey, A.~Jaworski, A.~Kaplan, H.~Khaitan,
  D.~Killebrew, A.~Koch, N.~Kumar, S.~Lacy, J.~Laudon, J.~Law, D.~Le, C.~Leary,
  Z.~Liu, K.~Lucke, A.~Lundin, G.~MacKean, A.~Maggiore, M.~Mahony, K.~Miller,
  R.~Nagarajan, R.~Narayanaswami, R.~Ni, K.~Nix, T.~Norrie, M.~Omernick,
  N.~Penukonda, A.~Phelps, J.~Ross, M.~Ross, A.~Salek, E.~Samadiani, C.~Severn,
  G.~Sizikov, M.~Snelham, J.~Souter, D.~Steinberg, A.~Swing, M.~Tan,
  G.~Thorson, B.~Tian, H.~Toma, E.~Tuttle, V.~Vasudevan, R.~Walter, W.~Wang,
  E.~Wilcox, and D.~H. Yoon, ``In-datacenter performance analysis of a tensor
  processing unit,'' in \emph{Proceedings of the 44th Annual International
  Symposium on Computer Architecture}, ser. ISCA '17.\hskip 1em plus 0.5em
  minus 0.4em\relax New York, NY, USA: ACM, 2017, pp. 1--12. [Online].
  Available: \url{http://doi.acm.org/10.1145/3079856.3080246}
\BIBentrySTDinterwordspacing

\bibitem{brainwave2018}
\BIBentryALTinterwordspacing
J.~Fowers, K.~Ovtcharov, M.~Papamichael, T.~Massengill, M.~Liu, D.~Lo,
  S.~Alkalay, M.~Haselman, L.~Adams, M.~Ghandi, S.~Heil, P.~Patel, A.~Sapek,
  G.~Weisz, L.~Woods, S.~Lanka, S.~K. Reinhardt, A.~M. Caulfield, E.~S. Chung,
  and D.~Burger, ``A configurable cloud-scale dnn processor for real-time ai,''
  in \emph{Proceedings of the 45th Annual International Symposium on Computer
  Architecture}, ser. ISCA '18.\hskip 1em plus 0.5em minus 0.4em\relax
  Piscataway, NJ, USA: IEEE Press, 2018, pp. 1--14. [Online]. Available:
  \url{https://doi.org/10.1109/ISCA.2018.00012}
\BIBentrySTDinterwordspacing

\bibitem{cudnn}
\BIBentryALTinterwordspacing
S.~Chetlur, C.~Woolley, P.~Vandermersch, J.~Cohen, J.~Tran, B.~Catanzaro, and
  E.~Shelhamer, ``cudnn: Efficient primitives for deep learning,'' \emph{CoRR},
  vol. abs/1410.0759, 2014. [Online]. Available:
  \url{http://arxiv.org/abs/1410.0759}
\BIBentrySTDinterwordspacing

\bibitem{tensorflow2016}
\BIBentryALTinterwordspacing
M.~Abadi, A.~Agarwal, P.~Barham, E.~Brevdo, Z.~Chen, C.~Citro, G.~S. Corrado,
  A.~Davis, J.~Dean, M.~Devin, S.~Ghemawat, I.~J. Goodfellow, A.~Harp,
  G.~Irving, M.~Isard, Y.~Jia, R.~J{\'{o}}zefowicz, L.~Kaiser, M.~Kudlur,
  J.~Levenberg, D.~Man{\'{e}}, R.~Monga, S.~Moore, D.~G. Murray, C.~Olah,
  M.~Schuster, J.~Shlens, B.~Steiner, I.~Sutskever, K.~Talwar, P.~A. Tucker,
  V.~Vanhoucke, V.~Vasudevan, F.~B. Vi{\'{e}}gas, O.~Vinyals, P.~Warden,
  M.~Wattenberg, M.~Wicke, Y.~Yu, and X.~Zheng, ``Tensorflow: Large-scale
  machine learning on heterogeneous distributed systems,'' \emph{CoRR}, vol.
  abs/1603.04467, 2016. [Online]. Available:
  \url{http://arxiv.org/abs/1603.04467}
\BIBentrySTDinterwordspacing

\bibitem{pytorch2017}
A.~Paszke, S.~Gross, S.~Chintala, G.~Chanan, E.~Yang, Z.~DeVito, Z.~Lin,
  A.~Desmaison, L.~Antiga, and A.~Lerer, ``Automatic differentiation in
  pytorch,'' in \emph{NIPS-W}, 2017.

\bibitem{gao2018cellular}
\BIBentryALTinterwordspacing
P.~Gao, L.~Yu, Y.~Wu, and J.~Li, ``Low latency rnn inference with cellular
  batching,'' in \emph{Proceedings of the Thirteenth EuroSys Conference}, ser.
  EuroSys '18.\hskip 1em plus 0.5em minus 0.4em\relax New York, NY, USA: ACM,
  2018, pp. 31:1--31:15. [Online]. Available:
  \url{http://doi.acm.org/10.1145/3190508.3190541}
\BIBentrySTDinterwordspacing

\bibitem{vinyalsTBE16}
\BIBentryALTinterwordspacing
O.~Vinyals, A.~Toshev, S.~Bengio, and D.~Erhan, ``Show and tell: Lessons
  learned from the 2015 {MSCOCO} image captioning challenge,'' \emph{CoRR},
  vol. abs/1609.06647, 2016. [Online]. Available:
  \url{http://arxiv.org/abs/1609.06647}
\BIBentrySTDinterwordspacing

\bibitem{miao2015eesen}
Y.~Miao, M.~Gowayyed, and F.~Metze, ``Eesen: End-to-end speech recognition
  using deep rnn models and wfst-based decoding,'' in \emph{Automatic Speech
  Recognition and Understanding (ASRU), 2015 IEEE Workshop on}.\hskip 1em plus
  0.5em minus 0.4em\relax IEEE, 2015, pp. 167--174.

\bibitem{hochreiter1997long}
S.~Hochreiter and J.~Schmidhuber, ``Long short-term memory,'' \emph{Neural
  computation}, vol.~9, no.~8, pp. 1735--1780, 1997.

\bibitem{cho14Gru}
\BIBentryALTinterwordspacing
K.~Cho, B.~van Merrienboer, {\c{C}}.~G{\"{u}}l{\c{c}}ehre, F.~Bougares,
  H.~Schwenk, and Y.~Bengio, ``Learning phrase representations using {RNN}
  encoder-decoder for statistical machine translation,'' \emph{CoRR}, vol.
  abs/1406.1078, 2014. [Online]. Available:
  \url{http://arxiv.org/abs/1406.1078}
\BIBentrySTDinterwordspacing

\bibitem{bucketing2016}
\BIBentryALTinterwordspacing
V.~Khomenko, O.~Shyshkov, O.~Radyvonenko, and K.~Bokhan, ``Accelerating
  recurrent neural network training using sequence bucketing and multi-gpu data
  parallelization,'' \emph{CoRR}, vol. abs/1708.05604, 2017. [Online].
  Available: \url{http://arxiv.org/abs/1708.05604}
\BIBentrySTDinterwordspacing

\bibitem{cheMxnet}
\BIBentryALTinterwordspacing
T.~Chen, M.~Li, Y.~Li, M.~Lin, N.~Wang, M.~Wang, T.~Xiao, B.~Xu, C.~Zhang, and
  Z.~Zhang, ``Mxnet: {A} flexible and efficient machine learning library for
  heterogeneous distributed systems,'' \emph{CoRR}, vol. abs/1512.01274, 2015.
  [Online]. Available: \url{http://arxiv.org/abs/1512.01274}
\BIBentrySTDinterwordspacing

\bibitem{wu2016google}
Y.~Wu, M.~Schuster, Z.~Chen, Q.~V. Le, M.~Norouzi, W.~Macherey, M.~Krikun,
  Y.~Cao, Q.~Gao, K.~Macherey \emph{et~al.}, ``Google's neural machine
  translation system: Bridging the gap between human and machine translation,''
  \emph{arXiv preprint arXiv:1609.08144}, 2016.

\bibitem{korf2009}
\BIBentryALTinterwordspacing
R.~E. Korf, ``Multi-way number partitioning,'' in \emph{Proceedings of the 21st
  International Jont Conference on Artifical Intelligence}, ser.
  IJCAI'09.\hskip 1em plus 0.5em minus 0.4em\relax San Francisco, CA, USA:
  Morgan Kaufmann Publishers Inc., 2009, pp. 538--543. [Online]. Available:
  \url{http://dl.acm.org/citation.cfm?id=1661445.1661531}
\BIBentrySTDinterwordspacing

\bibitem{rieraCGPA2019}
M.~{Riera}, J.~{Arnau}, and A.~{González}, ``Cgpa: Coarse-grained pruning of
  activations for energy-efficient rnn inference,'' \emph{IEEE Micro}, vol.~39,
  no.~5, pp. 36--45, Sep. 2019.

\bibitem{samajdar2018scale}
A.~Samajdar, Y.~Zhu, P.~Whatmough, M.~Mattina, and T.~Krishna, ``Scale-sim:
  Systolic cnn accelerator simulator,'' \emph{arXiv preprint arXiv:1811.02883},
  2018.

\bibitem{muralimanohar2009cacti}
N.~Muralimanohar, R.~Balasubramonian, and N.~P. Jouppi, ``Cacti 6.0: A tool to
  model large caches,'' \emph{HP Laboratories}, pp. 22--31, 2009.

\bibitem{micron}
{M}icron {I}nc., ``{TN}-53-01: {LPDDR}4 {S}ystem {P}ower {C}alculator,''
  \url{https://www.micron.com/support/tools-and-utilities/power-calc}.

\bibitem{yazdani2019lstmsharp}
R.~Yazdani, O.~Ruwase, M.~Zhang, Y.~He, J.-M. Arnau, and A.~Gonz\'{a}lez,
  ``Lstm-sharp: An adaptable, energy-efficient hardware accelerator for long
  short-term memory,'' 2019.

\bibitem{guan2017fpga}
Y.~Guan, Z.~Yuan, G.~Sun, and J.~Cong, ``Fpga-based accelerator for long
  short-term memory recurrent neural networks,'' in \emph{Design Automation
  Conference (ASP-DAC), 2017 22nd Asia and South Pacific}.\hskip 1em plus 0.5em
  minus 0.4em\relax IEEE, 2017, pp. 629--634.

\bibitem{deepcpu}
\BIBentryALTinterwordspacing
M.~Zhang, S.~Rajbhandari, W.~Wang, and Y.~He, ``Deepcpu: Serving rnn-based deep
  learning models 10x faster,'' in \emph{Proceedings of the 2018 USENIX
  Conference on Usenix Annual Technical Conference}, ser. USENIX ATC '18.\hskip
  1em plus 0.5em minus 0.4em\relax Berkeley, CA, USA: USENIX Association, 2018,
  pp. 951--965. [Online]. Available:
  \url{http://dl.acm.org/citation.cfm?id=3277355.3277446}
\BIBentrySTDinterwordspacing

\bibitem{grnn2019}
\BIBentryALTinterwordspacing
C.~Holmes, D.~Mawhirter, Y.~He, F.~Yan, and B.~Wu, ``Grnn: Low-latency and
  scalable rnn inference on gpus,'' in \emph{Proceedings of the Fourteenth
  EuroSys Conference 2019}, ser. EuroSys '19.\hskip 1em plus 0.5em minus
  0.4em\relax New York, NY, USA: ACM, 2019, pp. 41:1--41:16. [Online].
  Available: \url{http://doi.acm.org/10.1145/3302424.3303949}
\BIBentrySTDinterwordspacing

\bibitem{HanESE2017:}
\BIBentryALTinterwordspacing
S.~Han, J.~Kang, H.~Mao, Y.~Hu, X.~Li, Y.~Li, D.~Xie, H.~Luo, S.~Yao, Y.~Wang,
  H.~Yang, and W.~B.~J. Dally, ``Ese: Efficient speech recognition engine with
  sparse lstm on fpga,'' in \emph{Proceedings of the 2017 ACM/SIGDA
  International Symposium on Field-Programmable Gate Arrays}, ser. FPGA
  '17.\hskip 1em plus 0.5em minus 0.4em\relax New York, NY, USA: ACM, 2017, pp.
  75--84. [Online]. Available: \url{http://doi.acm.org/10.1145/3020078.3021745}
\BIBentrySTDinterwordspacing

\bibitem{masr2019}
U.~{Gupta}, B.~{Reagen}, L.~{Pentecost}, M.~{Donato}, T.~{Tambe}, A.~M. {Rush},
  G.~{Wei}, and D.~{Brooks}, ``Masr: A modular accelerator for sparse rnns,''
  in \emph{2019 28th International Conference on Parallel Architectures and
  Compilation Techniques (PACT)}, Sep. 2019, pp. 1--14.

\bibitem{riera2018computation}
M.~Riera, J.-M. Arnau, and A.~Gonz{\'a}lez, ``Computation reuse in dnns by
  exploiting input similarity,'' in \emph{Proceedings of the 45th Annual
  International Symposium on Computer Architecture}.\hskip 1em plus 0.5em minus
  0.4em\relax IEEE Press, 2018, pp. 57--68.

\bibitem{fuzzy2019}
\BIBentryALTinterwordspacing
F.~Silfa, G.~Dot, J.-M. Arnau, and A.~Gonz\'{a}lez, ``Neuron-level fuzzy
  memoization in rnns,'' in \emph{Proceedings of the 52Nd Annual IEEE/ACM
  International Symposium on Microarchitecture}, ser. MICRO '52.\hskip 1em plus
  0.5em minus 0.4em\relax New York, NY, USA: ACM, 2019, pp. 782--793. [Online].
  Available: \url{http://doi.acm.org/10.1145/3352460.3358309}
\BIBentrySTDinterwordspacing

\bibitem{paszke2017automatic}
A.~Paszke, S.~Gross, S.~Chintala, G.~Chanan, E.~Yang, Z.~DeVito, Z.~Lin,
  A.~Desmaison, L.~Antiga, and A.~Lerer, ``Automatic differentiation in
  {PyTorch},'' in \emph{NIPS Autodiff Workshop}, 2017.

\bibitem{seideCntk2016}
\BIBentryALTinterwordspacing
F.~Seide and A.~Agarwal, ``Cntk: Microsoft's open-source deep-learning
  toolkit,'' in \emph{Proceedings of the 22Nd ACM SIGKDD International
  Conference on Knowledge Discovery and Data Mining}, ser. KDD '16.\hskip 1em
  plus 0.5em minus 0.4em\relax New York, NY, USA: ACM, 2016, pp. 2135--2135.
  [Online]. Available: \url{http://doi.acm.org/10.1145/2939672.2945397}
\BIBentrySTDinterwordspacing

\end{thebibliography}

\end{document}